\documentclass[11pt]{article}
\usepackage{jheppubmod}
\pdfoutput=1
\usepackage{collref}
\usepackage{psfrag}
\usepackage{array}
\usepackage{amssymb}
\usepackage{amsmath}
\usepackage{amsthm}
\usepackage{graphicx}
\usepackage[labelsep=quad]{subcaption}
\usepackage{epstopdf}
	
\usepackage{epsfig}
\usepackage{tikz}
\usetikzlibrary{arrows,plotmarks,positioning,calc}

\newcommand{\hc}{\text{h.c}}

\newcommand{\rtor}{r_*}

\def\wp{\omega_{+}}
\def\wm{\omega_{-}}
\def\twoptnorm{{\cal N}}
\def\rplus{r_{+}}
\def\rmin{r_{-}}

\def\rconst{{\cal \zeta}}
\def\boga{{\cal \kappa}_1}
\def\bogta{\widetilde{\cal \kappa}_2}
\def\bogc{\widetilde{\cal \kappa}_3}
\def\bogd{{\cal \kappa}_4}
\def\boge{{\cal \kappa}_5}
\def\bogf{{\cal \kappa}_6}
\def\uent{\mathcal{U}}
\def\vent{\mathcal{V}}
\def\xent{\mathcal{\xi}}
\def\dxent{\delta \vec{\xent}}
\def\tune{{\cal T}}
\def\vol{ V_{\text{ol}}}

\def\op{{\cal O}}

\def\tb{\widetilde{b}}

\def\pb[#1,#2]{\{#1, #2\}}
\def\deb[#1,#2]{[#1,#2]_{\text{D.B.}}}

\def\hc{\text{h.c.}}

\def\a{\alpha}

\def\ta{\widetilde{a}}

\def\Or[#1]{{\text{O}}\left({#1}\right)}
\def\dotl[#1,#2]{\left\langle #1,\, #2 \right\rangle}
\def\dotlb[#1,#2]{\left\langle #1,\, #2 \right\rangle}
\def\dotlm[#1,#2]{\left[ #1,\, #2 \right]}
\def\dotp[#1,#2]{(\vect{#1} \cdot\vect{#2})}
\def\aff[#1,#2]{\hat{#1}(#2)}

\def\n4sym{{\cal N}=4 SYM}
\def\>{\rangle}
\def\<{\langle}
\def\weight[#1,#2,#3]{\{(#1),#2,#3\}}
\def\ads[#1]{$\text{AdS}_{#1}$}

\newcommand{\be}{\begin{equation}}
\newcommand{\ee}{\end{equation}}
\newcommand{\ba}{\begin{align}}
\newcommand{\ea}{\end{align}}
\newcommand{\bs}{\begin{split}}
\def\sess\end{split}
\newcommand{\vect}[1]{{\boldsymbol{#1}}}

\def\eq#1{{Eq.~(\ref{#1})}}
\def \bea {\begin{eqnarray}}
\def \eea {\end{eqnarray}}
\def \bea* {\begin{eqnarray*}}
\def \eea* {\end{eqnarray*}}

\def \bes {\begin{equation*}}
\def \ees {\end{equation*}}

\def \anh {\mathfrak{a}}
\def \sharp{\mathfrak{s}}
\def \dnh {\mathfrak{d}}
\def \enh {\mathfrak{e}}
\def \tildanh {\widetilde{\mathfrak{a}}}

\def \bnh {\mathfrak{b}}

\def \cnh {\mathfrak{c}}

\title{A simple quantum test for smooth horizons}
\author[a]{Kyriakos Papadodimas,}
\author[b]{Suvrat Raju}
\author[b]{and Pushkal Shrivastava}
\affiliation[a]{International Centre for Theoretical Physics,
Strada Costiera 11, 34151 Trieste, Italy}
\affiliation[b]{International Centre for Theoretical Sciences, Tata Institute of Fundamental Research, Shivakote, Bengaluru 560089, India.}
\emailAdd{kyriakos@ictp.it}
\emailAdd{suvrat@icts.res.in}
\emailAdd{pushkal.shrivastava@icts.res.in}
\date{}

\abstract{We develop a new test that provides a necessary condition for a quantum state to be smooth in the vicinity of a null surface: ``near-horizon modes'' that can be defined locally near any patch of the null surface must be correctly entangled with each other and with their counterparts across the surface. This test is considerably simpler to implement than a full computation of the renormalized stress-energy tensor. We apply this test to Reissner-Nordstr\"{o}m black holes in asymptotically anti-de Sitter space and provide numerical evidence that the inner horizon of such black holes is singular in the Hartle-Hawking state.  We then consider BTZ black holes, where we show that our criterion for smoothness is satisfied as one approaches the inner horizon from outside. This results from a remarkable conspiracy between the properties of mode-functions outside the outer horizon and between the inner and outer horizon.  Moreover, we consider the extension of spacetime across the inner horizon of BTZ black holes and show that it is possible to define modes behind the inner horizon that are correctly entangled with modes in front of the inner horizon. Although this provides additional  suggestions for the failure of strong cosmic censorship, we lay out several puzzles that must be resolved before concluding that the inner horizon will be traversable.}

\begin{document}
\maketitle

\section{Introduction \label{secintro}}
In addition to an outer event horizon---which characterizes a black hole---charged and rotating black holes may have additional horizons in their interior. These inner horizons are of theoretical interest, because they can also be Cauchy horizons.  Cauchy horizons present a problem for Laplacian determinism, because given smooth initial data in the past there is no unique way to evolve it past such a horizon even though the metric is not singular there.

This difficulty is sometimes bypassed by appealing to ``strong cosmic censorship''. This is the idea that although the Cauchy horizon exists for the static solution, the moment one considers generic initial data leading to the formation of a realistic black hole, the inner horizon will collapse to form a singularity and mark the end of spacetime.  The simplest physical intuition for this conjecture is as follows. In the static solution, the Cauchy horizon is to the future of all events outside the black hole. On the other hand, an infalling observer can reach the Cauchy horizon in a finite amount of proper time after crossing the outer horizon. Therefore, if one perturbs the exterior of the black-hole solution, then in the finite amount of time that it takes the observer to reach the inner horizon, the observer will receive radiation from all the events outside the black hole, which stretch out for an infinite amount of time \cite{penrose1968structure}. The strong cosmic censorship conjecture is based on the intuition that this infinite ``blue shift'' effect  will destabilize the inner horizon \cite{Simpson:1973ua,mcnamara1978instability,chandrasekhar1982crossing,Poisson:1989zz,Poisson:1990eh} . 

However, it is not straightforward to prove such a result \cite{Ori:1991zz,Ori:1992zz,Mellor:1989ac,Chambers:1994ap,Brady:1996za,Dafermos:2003wr,Dafermos:2002ka, Murata:2013daa,christodoulou2012formation,Bhattacharjee:2016zof}. Somewhat surprisingly,  in several cases,  it {\em is} possible to extend the metric continuously across the horizon \cite{Dafermos:2012np,Dafermos:2017dbw} and only the derivatives of the metric diverge. The classical analysis has therefore focused on the severity of this divergence and the question of whether solutions to the equations of motion can be continued weakly past the horizon. The answer to this question depends on the precise smoothness conditions that are placed on the initial data \cite{dafermos2018rough}.  One key physical issue in this analysis is a competition of the decay of perturbations outside the horizon, controlled by the quasinormal frequencies of the black hole, and the blue shift effect near the inner horizon controlled by its surface gravity \cite{Cardoso:2017soq}.  This has been the subject of extensive recent discussions \cite{Hod:2018dpx,Dias:2018ufh,Mo:2018nnu,Hod:2018lmi,Luk:2017jxq,Luna:2018jfk,Dias:2018ynt}  and we refer the reader to the introduction of \cite{Dias:2018etb} for a clear and concise discussion of the issues involved in the classical problem.

However, the world is not classical. So what happens when quantum effects are taken into account? This issue has received far less attention than the classical question. Moreover, the literature so far has focused on the behaviour of the renormalized stress-energy tensor \cite{hiscock1977stress,birrell1978falling, Dias:2019ery, Sela:2018xko, Steif:1993zv}  near the inner horizon. But this calculation is not  only tedious, it also suffers from inherent ambiguities \cite{wald1994quantum}.  

In this paper, we will propose a simple quantum-mechanical criterion that must be satisfied for the horizon to be smooth. Our test is much simpler to implement than a full computation of the stress-energy tensor, although we emphasize that it provides only a {\em necessary} and not a sufficient condition for the horizon to be smooth.

The test is as follows. Consider a scalar field, $\phi$ propagating in a background metric in $d+1$ dimensions. To investigate the smoothness of a null surface we consider two spacetime points $1$ and $2$ in its vicinity; the points may be on the same side of the surface or on opposite sides.  Then we demand that it should be possible to define a state of the quantum fields so that the short distance behaviour of the two-point function is given by
\be
\label{introshort}
\langle   \phi(1) \phi(2) \rangle = \twoptnorm  {1 \over s^{d-1 \over 2}} + \ldots,
\ee
where $s$ is the square of the geodesic distance between the two points and $\ldots$ denotes terms that are less-singular than the displayed term. (The normalization $\twoptnorm$ is chosen for convenience and described in section \ref{entanglednull}.)

At first sight, the condition \eqref{introshort} may appear trivial. After all, this is the term that is thrown away in computations of the stress-tensor. However, on further examination, the simple condition \eqref{introshort} turns out to yield a surprising number of riches. We first describe the physical intuition that suggests that condition \eqref{introshort} should be important, and then describe its mathematical consequences.

The physical intuition behind our test is that a smooth state in any spacetime is characterized by the existence of entangled degrees of freedom across any null surface. This is what allows spacetime to form a ``universal quantum channel''; the large degree of entanglement makes it possible to send arbitrary kinds of communications between one region and a nearby region. In fact, in any quantum field theory, a finite-energy state has so much entanglement between the various local degrees of freedom that by manipulating the degrees of freedom in any region of spacetime, it is possible to create arbitrary states in the Hilbert space \cite{Haag:1992hx}. The leading short-distance behaviour of the two-point function displayed in \eqref{introshort} is one quantification of the large amount of entanglement between neighbouring regions.

To distil this entanglement, we define appropriate near-horizon modes by integrating the field in local Rindler coordinates on both sides of the surface. Since the Rindler modes oscillate an infinite number of times near the surface, we can extract a sharp mode even by restricting ourselves to a very small region near the null surface.  These near-horizon modes depend on a frequency, $\omega_0$  and we show that if a state of the quantum fields, $|\Psi \rangle$ satisfies \eqref{introshort} then the action of these modes must be related through
\be
\label{acrosshorentintro}
(\anh - e^{-\pi \omega_0} \tildanh^{\dagger})|\Psi \rangle = 0,
\ee
and their two-point functions must be given by 
\be 
\label{samesideentintro}
\langle \Psi | \anh \anh^{\dagger} | \Psi \rangle = \langle \Psi | \tildanh \tildanh^{\dagger} | \Psi \rangle = {1 \over 1 - e^{-2 \pi \omega_0}}.
\ee

The reader will recognize that the Minkowski vacuum is annihilated by a similar combination of left and right Rindler modes \cite{birrell1984quantum}. The result above is slightly different since it uses only {\em local physics}; relatedly, the modes that appear are not defined by integrating the field over the half-space, but just in the neighbourhood of a null surface. In particular,  the relations \eqref{acrosshorentintro} and \eqref{samesideentintro} may fail to hold for globally defined  Rindler modes in an excited state even if it has finite Minkowski energy. On the other hand, for the near-horizon modes, \eqref{acrosshorentintro} and \eqref{samesideentintro} must hold in any smooth state. 

The significance of \eqref{acrosshorentintro} and \eqref{samesideentintro} can be brought out by considering a charged spherically symmetric black hole. In such a situation, the near-horizon modes reduce to globally defined Schwarzschild modes smeared with a function that is sharply peaked in frequency space. So the two-point function of the near-horizon modes picks up only the part of the correlators of the Schwarzschild modes  that is proportional to a delta function in the frequency-difference. In this simple setting, our result therefore implies the following: {\em for the horizon to be smooth in a given quantum state,  the two-point function of Schwarzschild modes must contain a piece proportional to a delta function in the frequency-difference with a coefficient that is precisely the Boltzmann factor corresponding to the temperature of the horizon.}

This criterion is powerful because it is not easy for a state to be smooth at both the outer and the inner horizons. For a state to be smooth at the outer horizon, the fields there must be entangled at the temperature corresponding to the outer horizon; then the propagation of fields between the inner and outer horizon must precisely ``heat'' (or ``cool'') them to the temperature of the inner horizon. This leads to an {\em infinite} sequence of constraints on the reflection and transmission coefficients that control the propagation of fields from the outer to the inner horizon.  Indeed, as we show in section \ref{secrnads}, a simple computation of these coefficients can be used to easily demonstrate that charged black holes in anti-de Sitter space have singular horizons without going through a full computation of the renormalized stress-tensor.

We can also apply our test to the rotating BTZ black hole. Here, as one might expect from the analysis of \cite{Dias:2019ery}, we find that our test is satisfied. This follows from a conspiracy between the properties of the reflection and transmission coefficients in the black-hole interior and the behaviour of normalizable modes outside the outer horizon. Since our test only provides a necessary condition for smoothness, it does not guarantee that the renormalized stress-energy tensor will be finite. Indeed, our test is satisfied even for those parameters of the BTZ black hole where the analysis of \cite{Dias:2019ery} suggests that the stress-tensor will diverge.

On the other hand, our test, in some respects, is also stronger than a check of the stress-tensor since it involves  modes on both sides of the horizon. In the case of the BTZ black hole, we are therefore forced to explore whether field operators can be extended smoothly past the inner horizon.

This extension turns out to be nontrivial since the standard construction of mirror operators \cite{Papadodimas:2013wnh,Papadodimas:2013jku} that applies to the outer horizon cannot be directly applied to the inner horizon of the BTZ black hole.  This can be understood in terms of the monogamy of entanglement. In the free-field limit,  the modes near the inner horizon can be written as linear-combinations of modes near the outer horizon. Since the modes near the outer horizon are entangled with modes outside the outer horizon, they cannot also be entangled with fresh-modes behind the inner horizon. We demonstrate a precise version of this result in the text.\footnote{We emphasize that this argument is very different, in flavour, from the monogamy arguments that were used in \cite{Mathur:2011uj,Almheiri:2012rt} since it can be implemented entirely at the level of effective field theory.} 

Surprisingly, it turns out to be possible to reuse the modes between the inner and outer horizon in such a manner that not only are the constraints of locality respected, the modes also have the correct two-point function dictated by  the temperature of the inner horizon. This also provides the {\em unique} extension of quantum fields in the near-horizon region just beyond the inner horizon if the inner horizon is traversable.

At first sight, our results provide additional evidence for the claim that strong cosmic censorship fails for the BTZ black hole. If so, this would lead to a very puzzling situation from the point of view of the AdS/CFT correspondence \cite{Maldacena:1997re}. If the inner horizon is traversable, an infalling observer in an eternal BTZ black hole can simply exit the system as described by the two CFTs. This would imply that the eternal BTZ black hole is not completely described by two entangled CFTs \cite{Maldacena:2001kr} and this system must be supplemented by additional degrees of freedom.

However, as we explain in section \ref{secdiscussion}, the question of  traversability of the inner horizon is more delicate. For instance, if one perturbs the exterior of the black hole, although the perturbation initially dies off at the quasinormal frequency,  unitarity requires a nonperturbative tail (of size $e^{-{S \over 2}}$) to survive at late times \cite{Maldacena:2001kr}. What is the effect of this tail on the inner horizon? Relatedly, the boundary CFT undergoes spontaneous fluctuations over exponential long time-scales. These fluctuations, which appear at arbitrarily late times, can also destroy the inner horizon. In this paper, we do not present a solution to these questions, but leave them open for further work.

The plan of the paper is as follows. In section \ref{entanglednull} we describe our test in greater detail, define the near-horizon modes carefully and derive \eqref{acrosshorentintro} and \eqref{samesideentintro} starting with \eqref{introshort}. We then apply this test to charged asymptotically AdS black holes in various dimension in section \ref{secrnads} and to the BTZ black hole in section \ref{secbtz}. We conclude in section \ref{secdiscussion}. The Appendix presents a WKB analysis of the propagation of large angular-momentum waves in AdS black holes, which is of interest since in this regime  \eqref{acrosshorentintro} and \eqref{samesideentintro} are automatically satisfied at the inner horizon if they are satisfied at the outer horizon.

\section{Entangled modes across a null surface \label{entanglednull}}
 In quantum field theory, to test whether the spacetime in the vicinity of a surface is smooth, we can test whether it is possible to transmit ``messages'' in that region by turning on a source for a field at one point and measuring the response of the field at another. While the long-distance propagation of the source will depend on the nature of the spacetime, in the short-distance limit, we expect that fields in the neighbourhood of the source will respond in a universal manner. The response of the one-point function of the field to a source depends only on the commutator, but the response of higher-point functions also requires a specific short-distance behaviour for the two-point function. This, in turn, requires degrees of freedom that constitute the quantum field to be entangled with each other in a specific manner, as we now explain.

\begin{figure}[h!]
\begin{center}
\includegraphics[width=0.3\textwidth]{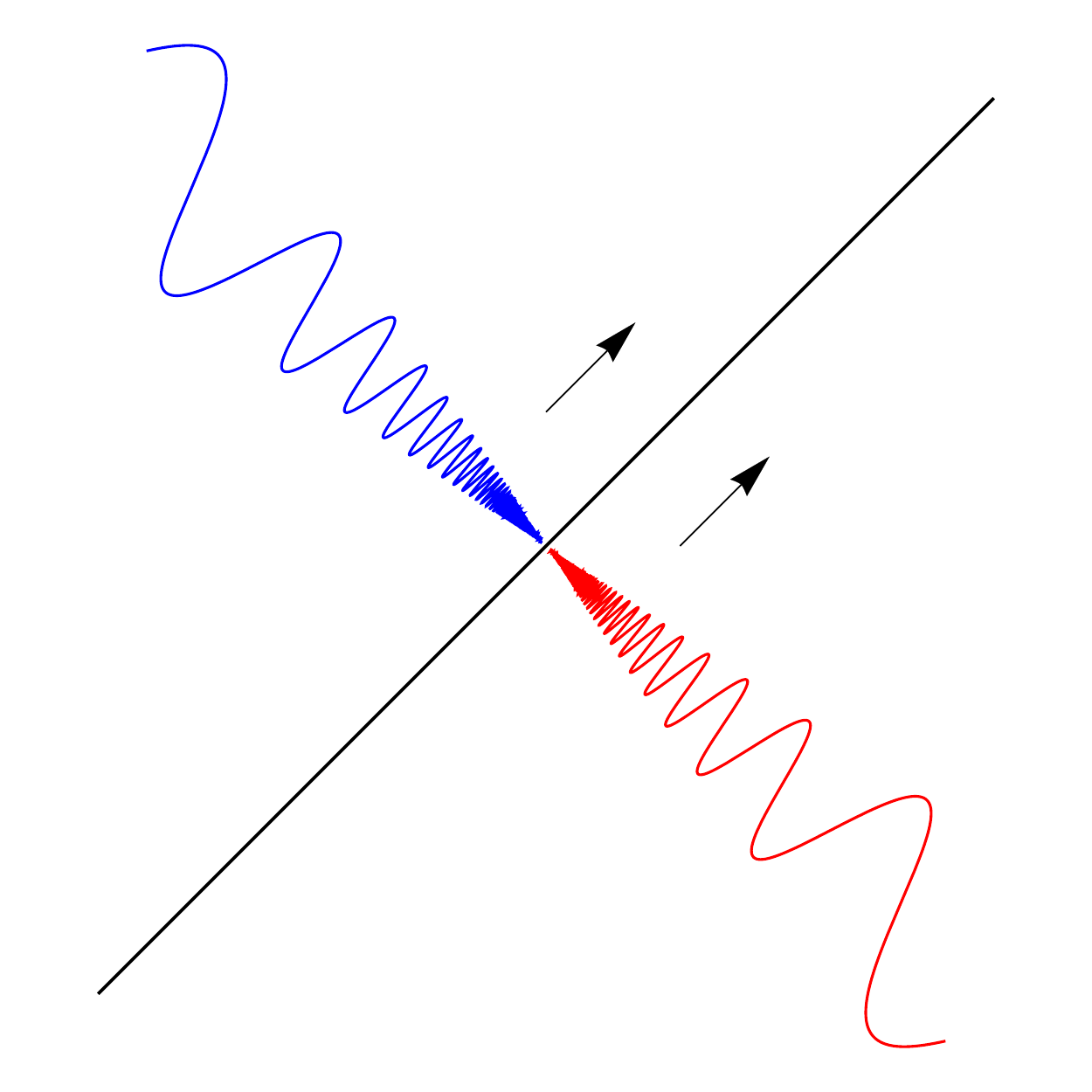}
\end{center}
\caption{\em Near horizon right-moving modes of a free scalar field near a null surface (black line). The red modes correspond to the operator $\anh$  while the blue ones to $\tildanh$ (see eq. \eqref{defnearhormodes}). The arrows indicate that the modes propagate parallel to the surface.}
\label{nearhorizonmodes}
\end{figure}

 In the neighbourhood of any null surface, the quantum fields can be expanded in a set of modes corresponding to left and right movers. We will choose the convention that the ``left movers'' are those that smoothly cross the surface, so that the phase factor multiplying these modes varies smoothly as we move along the surface. On the other hand, the modes that move parallel to the null surface (i.e. whose surfaces of constant phase are parallel to the null surface)  will be called ``right movers''. See Figure \ref{nearhorizonmodes}.

The specific technical result that we intend to show is that it is possible to define appropriate ``right moving'' modes on the two sides of the null surface, so that their action on the state of the system is related in a specific manner.\footnote{We use the term ``entanglement'' to denote this relationship although our result is primarily about correlation functions, and not about any quantum-information measure of entanglement. The reason for our terminology is that --- as shown in \cite{Raju:2018zpn} --- it is possible to use these correlators between near-horizon modes to extract Bell pairs from opposite sides of the horizon.}

We consider a small portion of a null surface, such that the metric is continuous across the surface. 
In $(d+1)$-dimensions, we introduce coordinates $\uent,\vent$ and transverse coordinates, $\xent^{a}$ with $2 \leq a \leq d$. We use a calligraphic font for these coordinates in this section to distinguish them from the coordinates that appear in the black-hole geometry.  The Greek indices below can take values in a larger range, $0 \leq \mu, \nu \leq d$, and we set $\xent^0 = \uent$ and $\xent^1 = \vent$. The null surface under consideration is that of $\uent = 0$, and on this surface, we will consider the small patch near $\xent^{\mu} = 0$, where we write the metric as 
\be
\label{nearpatchmetric}
ds^2 = -d \uent d \vent + \delta_{\alpha \beta} d \xent^{\alpha} d \xent^{\beta}  + g_{\mu \nu}^{(1)} d\xi^{\mu} d \xent^{\nu}.
\ee
We will assume that $g_{\mu \nu}^{(1)} \rightarrow 0$ as $\xent^{\mu} \rightarrow 0$ i.e. as we approach the patch under consideration the metric is well approximated by the first two terms in \eqref{nearpatchmetric}.

\paragraph{ Near-horizon 2-point function\\}

Now consider a scalar field propagating in this background. The precise assumption that we will make is that if the spacetime is ``reasonable'' then, in the short-distance limit
the two-point function of the scalar field will be given simply by an inverse power of the geodesic distance. This is true at least if the ultraviolet physics of the scalar operator is controlled by a free-fixed point (at scales much lower than the Planck scale). A similar assumption can be made for fermions, or fields with spin. 

Consider two insertions of the field at two nearby, but distinct points, which may be on the same side or on opposite sides of the null surface. We denote the state of the full system by $| \Psi \rangle$. In AdS/CFT, this state may be considered to be a state in the Hilbert space, as defined by the boundary conformal field theory.  The assumption above implies that if this state is ``reasonable'' then
\be
\label{twoptfn}
\langle \Psi | \phi(\uent_1, \vent_1, \xent_1^{a}) \phi(\uent_2, \vent_2, \xent_2^{a}) | \Psi \rangle = \twoptnorm  {1 \over s^{d-1 \over 2}} + {\cal R}.
\ee
In the equation above, the leading-order square of the geodesic distance $s$  is given by
\be
\label{geodist}
s = (-\delta \uent \,\delta \vent  + \delta_{\alpha \beta} \delta \xent^{a} \delta \xent^{b}),
\ee
where $\delta \uent = \uent_1 - \uent_2 - i \epsilon; \delta \vent = \vent_1 - \vent_2 - i \epsilon$ and we use the $(d-1)$-vector, $\delta \vec{\xent}$, with components $\delta \vec{\xent}^{a} = \xent_1^{a} - \xent_2^{a}$ to denote the displacement in the transverse directions between the two points. Here,   ${\cal R}$ can be any function of the variables $(\uent_i, \vent_i,\xent_i^a)$ that is less-singular than the displayed leading term as the points come close to each other. We have also introduced $-i \epsilon$ regulators with $\delta \uent$ and $\delta \vent$, which reflect that fact that we are interested in a Wightman function. This will be important when we consider commutators below. The normalization of the short-distance singularity
\be
\twoptnorm= {\Gamma(d-1) \over 2^d \pi^{d \over2} \Gamma({d \over 2})},
\ee
 can be fixed by considering a canonically normalized field in flat-space and explicitly checking its short-distance behaviour. 

The important equation \eqref{twoptfn} makes the assumptions that the short-distance singularities of the two-point function arise  only when the geodesic distance vanishes.  Note that the remaining part of the metric $g^{(1)}_{\mu \nu}$ does not appear in the leading singular part of this expression, although it is important for the subleading terms captured in the function ${\cal R}$. 

We will show that \eqref{twoptfn} forces a specific form of entanglement between right movers across the surface $\uent=0$. 
To see this, we take a further limit of \eqref{twoptfn} and repeat the manipulations that led to equation (4.3) of \cite{Papadodimas:2015jra}. If we first differentiate the two-point function, we will find that
\be
\langle \Psi | \partial_{\uent_1} \phi(\uent_1, \vent_1,\xent_1^{a}) \partial_{\uent_2} \phi(\uent_2, \vent_2, \xent_2^{a}) | \Psi \rangle =  -{d^2 - 1 \over 4 }  {\twoptnorm }  {(\delta \vent)^2 \over s^{{d+3 \over 2}}} + \partial_{\uent_1} \partial_{\uent_2} {\cal R}.
\ee
We now consider the limit $\delta \vent \rightarrow 0$. In this limit the derivative of the two-point function above  goes to zero unless $\dxent = 0$. But this point, where the transverse displacement vanishes, gives a delta function contribution. We can check this, and also determine the normalization of the delta function by performing an integral over the transverse coordinates
\be
\begin{split}
\lim_{\delta \vent \rightarrow 0} \int d^{d-1} \dxent  {(\delta \vent)^2 \over s^{{d+3 \over 2}}} &= \lim_{\delta \vent \rightarrow 0} {1 \over (\delta \uent)^2} {1 \over (-\delta \uent \delta \vent)^{{d-1 \over 2}}} \int d^{d-1} \dxent \left(1 + {1 \over (-\delta \uent \delta \vent)} \dxent^2 \right)^{-{d+3\over 2}}\\  &= {1 \over (\delta \uent)^2} \kappa_N,
\end{split}
\ee
where
\be
\kappa_N = \int d^{d-1} \dxent {1 \over (1 + \delta \vec{\xent}^2)^{d + 3 \over 2}} = {4 \over d^2 - 1}  {\pi^{d - 1 \over  2} \over \Gamma({d - 1 \over 2})}.
\ee
It is important that, in the integral above, we take the range of the integral over the transverse coordinates to be $(-\infty, \infty)$ in order to be able to change variables from $\dxent \rightarrow  {\dxent \over \sqrt{-\delta \uent \delta \vent}}$. This is only a trick to determine the normalization. It does {\em not} require that the expression for the two-point function remain valid for an infinite separation in transverse coordinates or even that the transverse coordinates have an infinite extent.

The various normalization constants in the equations above simplify when taken together using 
\be
{d^2 - 1 \over 4} \kappa_N \twoptnorm = {1 \over 4 \pi}.
\ee

Note that if we were to perform the operations above on a term in the two-point function that diverges with a smaller power of geodesic distance or on a regular function, which may appear inside ${\cal R}$, we would obtain a result that is  non-singular in the limit $\uent_1 \rightarrow \uent_2$. 

So  we finally find
\be
\begin{split}
\label{shortdistlimittwopt}
\lim_{\vent_1 - \vent_2 \rightarrow 0} \langle \Psi | \partial_{\uent_1} \phi(\uent_1, \vent_1, \xent_1^a) \partial_{\uent_2} \phi(\uent_2, \vent_2, \xent_2^a) | \Psi \rangle &= -{1 \over 4 \pi} {1 \over (\uent_1 - \uent_2 - i \epsilon)^2} \delta^{d-1}(\dxent)\\
&+ \lim_{\vent_1 - \vent_2 \rightarrow 0} \partial_{\uent_1} \partial_{\uent_2} \tilde{R}.
\end{split}
\ee
If the reader is worried about the ultralocal delta function that appears above, she should note that in the applications below we will always use \eqref{shortdistlimittwopt} after integrating both sides with test functions. Typically, we will consider fields that are separated by $\epsilon_{\vent}$ in the $\vent$ coordinate, and smeared over a range $\epsilon_{\uent}$ in the $\uent$ coordinate and $\epsilon_{\xent}$ in the transverse coordinates. Then the formula above holds provided we take $\epsilon_{\vent} \ll \epsilon_{\uent} \ll \epsilon_{\xent}$. 

\paragraph{ Near-horizon modes\\}

We now define some modes by integrating the field over a very short distance on both sides of the $\uent=0$ surface and for a very short distance over the transverse coordinates. We will show that the short distance structure of the two-point function above forces these modes to have a specific two-point correlation function. We follow the procedure given in \cite{Raju:2018zpn} to define the modes.   

The intuition is the following: as we approach the null surface from below  $\uent\rightarrow 0^-$ we can find solutions of the wave equation that behave like $(-\uent)^{i \omega_0}$. These solutions undergo an infinite number of oscillations until $\uent=0$. If we think in terms of the tortoise-like coordinate $u=\log(-\uent)$ these solutions behave like plane waves. This approximation becomes better as we approach the null surface at $\uent\rightarrow 0$ or $u=-\infty$. We define modes $\anh$ corresponding to wave packets with highly peaked frequency $\omega_0$ and centered very far towards $u=-\infty$, or equivalently around a point $\uent_0$ very near the null surface $\uent=0$. We make sure that these modes are unit-normalized (as opposed to having delta-function normalization). We define these modes as Fourier modes with approximate frequency $\omega_0$ characterized by a window of support in position space centered around $\uent_0$ and with a very large width in $u$-coordinates (though a small region in $\uent$ coordinates). 

This is achieved by introducing a ``tuning function'' $\tune(\uent)$ which has the property that $\tune(\uent)$ is real and has support only for $\uent \in [\uent_l, \uent_h]$, where $0<\uent_l \ll \uent_0 \ll\uent_h$ and so that $\uent_h$ is much smaller than the characteristic curvature scale of the geometry. 
This tuning function is assumed to vanish smoothly at the end-points of its support and normalized carefully as follows. First, we define its Fourier transform via
\be
\label{sharpdef}
\tune(\uent)   = \int_{-\infty}^{\infty} \sharp(\nu) \left({\uent \over \uent_0} \right)^{i \nu} d \nu; \qquad \sharp(\nu)   = {1 \over 2 \pi} \int_{0}^{\infty} {d \uent \over \uent} \tune(\uent) \left({\uent \over \uent_0} \right)^{-i \nu}. 
\ee
Then we normalize the ``tuning function'' by demanding
\be
\int   |\sharp(\nu)|^2 {d \nu} = 1.
\label{normtuning}
\ee
We choose $\sharp(\nu)$ to be  sharply peaked around $\nu = 0$, which corresponds to $\tune(\uent)$ being almost constant in the domain $[\uent_l,\uent_h]$.

We now also integrate over a small volume $\vol$ in the transverse coordinates and define the modes
\be
\label{defnearhormodes}
\begin{split}
&\anh = {1 \over  \sqrt{\pi \omega_0}  }\int  {d^{d-1} \xent^a \over \sqrt{\vol}} \int_0^{-\infty} d \uent \partial_{\uent} \phi(\uent, \vent=0, \xent^a) \left({-\uent \over \uent_0} \right)^{-i \omega_0} \tune(-\uent), \\
&\tildanh = {1 \over  \sqrt{\pi \omega_0}  }\int {d^{d-1} \xent^a \over \sqrt{\vol}} \int_0^{\infty} \partial_\uent \phi(\uent,\vent=-\epsilon,\xent^a) \left({\uent \over \uent_0} \right)^{i \omega_0} \tune(\uent).
\end{split}
\ee
The domain of integration, which is controlled by $\tune(\uent)$ is a very small region on both sides of the null surface and also a very small patch in the transverse coordinates.  
The modes have a nontrivial dependence on the choice of $\omega_0$  and, moreover, such modes can be defined in the vicinity of any point of the null surface.  But we have suppressed this dependence on the left hand sides of \eqref{defnearhormodes} to lighten the notation. 

In the rest of this section we will show how the short-distance behaviour of the two-point function \eqref{shortdistlimittwopt} determines the two-point correlators of these modes. 

Before moving on, we emphasize that the modes $\anh,\tildanh$ depend on the choice of $\uent_0,\uent_l,\uent_h$ and also the precise choice of the tuning function. We are interested in the behavior of these modes in the limit $\uent_0\rightarrow 0$ and ${\uent_h\over \uent_0},{\uent_0\over \uent_l}\gg1$ and where $\sharp(\nu)$ is sharply peaked. The statements we are going to make about the modes $\anh,\tildanh$ correspond to the behavior of these modes in this particular limit. However, in the equations below,  we will not display the dependence on these various cutoffs explicitly,  since the equations  can be made arbitrarily precise by taking these cutoffs to be as small as necessary. 

\paragraph{ Commutators\\}
If we assume that the field operators satisfy canonical commutation relations then we find that  at equal values of $\vent$ we have
\be
[\phi(\uent_1, \vent, \xent_1^a), \partial_{\uent_2} \phi(\uent_2, \vent, \xent_2^a)] = {i \over 2} \delta^{d-1}(\dxent) \delta(\uent_1 - \uent_2).
\label{lightconeccr}
\ee
This commutators has, inbuilt into it, the fact that the fields commute at spacelike separation. We note that the commutator is also consistent with the two-point correlator that we found above because
\be
\text{Im} \left(\langle  \partial_{\uent_1} \phi(\uent_1, \vent, \xent_2^a), \partial_{\uent_2}  \phi(\uent_2, \vent, \xent_1^a) \rangle \right) = \text{Im}\left[{-{1 \over 4 \pi}} {1 \over (\uent_1 - \uent_2 - i \epsilon)^2}  \right] = {1 \over 4} \delta'(\uent_1 - \uent_2) \delta^{d-1}(\dxent),
\ee
which can be seen from the identity $-{1 \over (x - i \epsilon)^2} = {\partial \over \partial x} {1 \over x - i \epsilon} = {\partial \over \partial x} {\cal P}{1 \over x} + i \pi \delta'(x)$. 

We can substitute these commutation relations into the definition of the modes to compute their commutators. 
\be
\begin{split}
[\anh, \anh^{\dagger}] =  {1 \over  \pi \omega_0 \vol } \int &\left[\partial_{\uent_1} \phi(\uent_1, \vent=0, \xent^a_1), \partial_{\uent_2} \phi(\uent_2, \vent=0, \xent^a_2) \right]  \\ &\times \left({-{\uent_1 \over \uent_0}} \right)^{-i \omega_0} \left({-{\uent_2 \over \uent_0}} \right)^{i \omega_0}  \tune(-\uent_1)  \tune(-\uent_2)  d \uent_1 d \uent_2 d^{d-1} \xent^a_1  d^{d-1} \xent^a_2.
\end{split}
\ee
Using the canonical commutators \eqref{lightconeccr}, and doing the trivial integral in the transverse directions, we find that
\be
\begin{split}
[\anh, \anh^{\dagger}] &=  {i \over  2 \pi \omega_0 } \int  \partial_{\uent_1} \delta (\uent_1 - \uent_2) \left({-{\uent_1 \over \uent_0}} \right)^{-i \omega_0} \left({-{\uent_2 \over \uent_0}} \right)^{i \omega_0}  \tune(-\uent_1)  \tune(-\uent_2)  d \uent_1 d \uent_2 \\
&= {i \over  2 \pi \omega_0 } \int_0^{-\infty}  \partial_{\uent_1} \tune(-\uent_1)  \tune(-\uent_1)  d \uent_1  + {1 \over 2 \pi } \int_0^{-\infty} \tune(-\uent_1)^2 {d \uent_1  \over \uent_1}.
\end{split}
\ee
In doing the integral over the delta function in the $\uent$ coordinates, we were careful that the $\uent_2$ integral proceeds with $d \uent_2 < 0$, and so the delta function sets $\uent_1 = \uent_2$ and gives an additional minus sign. 
The first term above vanishes if the tuning function vanishes at both its end points. For the second term, using the normalization of the tuning function above we find
\be
[\anh, \anh^{\dagger}] = 1.
\ee
A similar calculation for $\tildanh$ leads to
\be
[\tildanh, \tildanh^{\dagger}] = 1,
\ee
and also
\be
[\anh, \tildanh] =[\anh, \tildanh^{\dagger}] = 0.
\ee

\paragraph{ Cross-correlators\\}
For the cross-correlators we have
\be
\begin{split}
\langle \Psi | \anh \tildanh | \Psi \rangle = {1 \over  \pi \vol \omega_0 } \int d \uent_1 d \uent_2  &\partial_{\uent_1}  \partial_{\uent_2} \langle \Psi | \phi(\uent_1, \vent=0, \xent_1^a) \phi(\uent_2, \vent=0, \xent_2^a) | \Psi \rangle \\ &\times (-\uent_1)^{-i \omega_0} (\uent_2)^{i \omega_0} \tune(-\uent_1) \tune(\uent_2) d^{d-1} \xent_1^a d^{d-1} \xent_2^a.
\end{split}
\ee
In the small domain where the integrand above has support,  note that the regular parts of the two-point function just drop out and the only contribution comes from the singular terms identified above.

Now, note that we may write
\be
{1 \over (\uent_1 - \uent_2)^2} = {1 \over (-\uent_1) \uent_2} \int_{-\infty}^{\infty} \omega {e^{-\pi \omega} \over 1 - e^{-2 \pi \omega}} \left(-{\uent_2 \over \uent_1} \right)^{-i \omega} d \omega,
\ee
where $\uent_1 < 0$ and $\uent_2 > 0$. 
If $|\uent_1| > |\uent_2|$ this identity follows from completing the $\omega$ integral in the upper  half plane and picking up the poles at $\omega = i n$; else we close the contour in the lower half-plane and pick up the poles at $\omega=-i n$. 

Substituting the short distance limit \eqref{shortdistlimittwopt} into the expression above and doing the transverse integral we find that
\be
\begin{split}
&\langle \Psi | \anh \tildanh | \Psi \rangle = {1 \over 4 \pi^2 \omega_0} \int {d \uent_1 \over \uent_1}  {d \uent_2 \over \uent_2} \omega  {e^{-\pi \omega} \over 1 - e^{-2 \pi \omega}} \left(-{\uent_2 \over \uent_1} \right)^{-i \omega} d \omega \tune(-\uent_1) \tune(\uent_2)  (-\uent_1)^{-i \omega_0} (\uent_2)^{i \omega_0}   \\
&= {1 \over \omega_0} \int \omega {e^{-\pi \omega} \over 1 - e^{-2 \pi \omega}} |\sharp(\omega - \omega_0)|^2 d \omega.
\end{split}
\ee
In the limit where $\sharp(\omega)$ is very sharply peaked around $\omega = 0$, and using the normalization condition \eqref{normtuning},  the two-point function just reduces to
\be
\label{universalentang}
\langle \Psi | \anh \tildanh | \Psi \rangle = {e^{-\pi \omega_0} \over 1 - e^{-2 \pi \omega_0}}.
\ee
This gives us the universal form of entanglement for short distance modes on opposite sides of the horizon. As in \cite{Raju:2018zpn}, this two-point function can be used to extract Bell pairs from either side of the null surface. 

Taking the Hermitian conjugate of this relation leads to
\be
\label{universalentangstar}
\langle \Psi | \anh^{\dagger} \tildanh^{\dagger} | \Psi \rangle = {e^{-\pi \omega_0} \over 1 - e^{-2 \pi \omega_0}}.
\ee
Finally, by a similar calculation we find
\be
\label{crosszero}
\langle \Psi| \anh^{\dagger} \tildanh |\Psi\rangle = \langle \Psi| \anh \tildanh^{\dagger} |\Psi\rangle  = 0.
\ee
\paragraph{ Self-correlators \\}
We can also determine correlators of the $\anh$ and $\tildanh$ modes with their own conjugates.  In this case, we need the identity 
\be
\label{iepidentity}
{-1 \over (\uent_1 - \uent_2 - i \epsilon)^2} = {1 \over  \uent_1 \uent_2} \int_{-\infty}^{\infty} \omega {1 \over 1 - e^{-2 \pi \omega}} \left({\uent_2 \over \uent_1} e^{-i \epsilon} \right)^{-i \omega} d \omega,
\ee
when $\uent_1 < 0$ and $\uent_2 < 0$. The integral on the right can again be done by closing the $\omega$ contour in the upper or the lower half plane and picking up the poles at $\omega = i n$. Note that the $i \epsilon$ term ensures that the integral converges both at large positive and large negative $\omega$. 

Using this identity,  we see that
\be
\label{universalself}
\begin{split}
&\langle \Psi | \anh \anh^{\dagger} | \Psi \rangle = \int {d \uent_1 d \uent_2 \over \uent_1 \uent_2} {1 \over 1 - e^{-2 \pi \omega}} \left({\uent_2 \over \uent_1} \right)^{-i \omega} d \omega \tune(-\uent_1) \tune(-\uent_2) \left({- {\uent_1 \over \uent_0 }}\right)^{-i \omega_0} \left({-{\uent_2 \over \uent_0}} \right)^{i \omega_0}  \\
&= \int {1 \over 1 - e^{-2 \pi \omega}} |\sharp(\omega - \omega_0)|^2 d \omega.
\end{split}
\ee
Again, in the limit where $\sharp(\omega)$ is sharply peaked we find that
\be
\label{selfcorrelator}
\langle \Psi | \anh \anh^{\dagger} | \Psi \rangle = 1 + \langle \Psi | \anh^{\dagger} \anh | \Psi \rangle = {1 \over 1 - e^{-2 \pi \omega_0}}.
\ee

A small subtlety that we note is that it is the $i \epsilon$ prescription that is important for ensuring that  $\langle \anh \anh^{\dagger} \rangle$ and $\langle \anh^{\dagger} \anh \rangle$ have distinct values. Indeed, the reader might have wondered whether it is possible to replace ${1 \over 1- e^{-2 \pi \omega}} \rightarrow {e^{-2 \pi \omega} \over 1 - e^{-2 \pi \omega}}$ in the identity \eqref{iepidentity} since both these terms have the same residues at the pole $\omega = \pm i n$. However, with an additional factor of $e^{-2 \pi \omega}$ in the numerator, the integral will not converge for large negative $\omega$ since the numerator and the denominator now grow at the same rate but the factor of $e^{-\epsilon \omega}$ grows in that regime. It is the $i \epsilon$ prescription that forces us to use the measure displayed in the equation \eqref{iepidentity} rather than the one with an additional exponential factor. This, in turn, ensures that there is no exponential factor in the numerator on the right hand side of \eqref{selfcorrelator}.

\paragraph{ Relating the action of modes on the state \\}
Above, we derived the two-point functions between the various modes. However, by putting these results together, we can derive a stronger relationship that relates the action of the left-movers on the quantum state to the action of the right movers.

Let us write 
\be
\tildanh |\Psi \rangle = c_1 \anh |\Psi \rangle + c_2 \anh^{\dagger} | \Psi \rangle + |\chi \rangle,
\ee
where $c_1, c_2$ are constants to be determined and $|\chi \rangle$ is orthogonal to the vectors produced by the action of $\anh$ and $\anh^{\dagger}$ on $|\Psi \rangle$  so that
\be
\langle \chi | \anh |\Psi \rangle = \langle \chi | \anh^{\dagger} | \Psi \rangle = 0.
\ee
Now, from \eqref{crosszero} we have 
\be
\langle \Psi | \anh^{\dagger} \tildanh | \Psi \rangle = 0 \Rightarrow c_1 = 0.
\ee
We can now set $c_2$ using
\be
\langle \Psi | \anh \tildanh | \Psi \rangle = {e^{-\pi \omega_0} \over 1 - e^{-2\pi \omega_0}} \Rightarrow   c_2 = e^{-\pi \omega_0}.
\ee
Finally, note that
\be
\langle \Psi | \tildanh^{\dagger} \tildanh | \Psi \rangle = {e^{-2 \pi \omega_0} \over 1 - e^{-2 \pi \omega_0}} \Rightarrow \langle \chi | \chi  \rangle = 0.
\ee
where we have also used the two-point correlator of $\anh$ with its conjugate. As similar procedure can be followed for the action of $\tildanh^{\dagger}$. Therefore, we reach the simple relations
\be
\label{universalrelatedaction}
\tildanh |\Psi \rangle = e^{-\pi \omega_0} \anh^{\dagger} |\Psi \rangle; \quad \tildanh^{\dagger} |\Psi \rangle = e^{\pi \omega_0} \anh | \Psi \rangle.
\ee

\paragraph{Null surfaces with spherical or translational symmetry\\}
The analysis above can be simplified slightly if the null surface under consideration has spherical or translational symmetry. Since the case of spherical symmetry is the case that will be important for us later, we indicate the generalization explicitly.

Consider a spherically symmetric null surface so that the metric is given by
\be
ds^2 = -d \uent d \vent + r_0^2 d \Omega_{d-1}^2 +  g_{\mu \nu}^{(1)} d\xent^{\mu} d \xent^{\nu}.
\ee
We now impose that $g_{\mu \nu}^{(1)} \rightarrow 0$ provided that $\uent, \vent \rightarrow 0$,  and, as above, Greek indices, $\mu, \nu$ run over all coordinates.

We can now define modes by integrating all over the $(d-1)$-sphere and extracting a particular spherical harmonic. 
\be
\label{sphericalmodedef}
\begin{split}
&\anh = {r_0^{{d-1 \over 2}}  \over  \sqrt{\pi \omega_0}  } \int\partial_{\uent} \phi(\uent, \vent=0, \Omega) \left({-\uent \over \uent_0} \right)^{-i \omega_0} \tune(-\uent) d \uent Y_{\ell}^*(\Omega) d^{d-1} \Omega, \\
&\tildanh = {r_0^{{d-1 \over 2}}  \over  \sqrt{\pi \omega_0}} \int \partial_\uent \phi(\uent,\vent=-\epsilon,\Omega) \left({\uent \over \uent_0} \right)^{i \omega_0} \tune(\uent) d \uent Y_{\ell}(\Omega) d^{d-1} \Omega,
\end{split}
\ee
Now, rather than being defined in the neighbourhood of a point, the modes carry an angular momentum, denoted by $\ell$.  These modes depend both on $\ell$ and on $\omega_0$ but we have suppressed that dependence to lighten the notation.

A precise repetition of the analysis above, leads to the same result for the action of these modes on the state
\be
\label{universalrelatedaction}
\tildanh |\Psi \rangle = e^{-\pi \omega_0} \anh^{\dagger} |\Psi \rangle; \quad \tildanh^{\dagger} |\Psi \rangle = e^{\pi \omega_0} \anh | \Psi \rangle.
\ee
Note that the orthogonality of the spherical harmonics implies that the modes are entangled only if the same value of $\ell$ is used in both lines of  \eqref{sphericalmodedef}.

\paragraph{ A note of caution \\}
The relation \eqref{universalrelatedaction} holds for the special modes that we have defined, which use information from the field just next to the null surface. In particular, even in the simple background of a Schwarzschild black hole, these modes must be distinguished from the global Schwarzschild modes that are defined by integrating the field all over the exterior or all over the interior.

The difference between these modes is manifest in their response to perturbations. If we consider perturbations of the horizon generated by just throwing in some matter on top of an equilibrium black hole then this will change the form of the two-point function between Schwarzschild modes in the interior and the exterior.
However, this perturbation does not affect the nature of the entanglement very close to the horizon. In particular, the entanglement between the modes $\anh$ and $\tildanh$, which  are defined by integrals  very  close to the null surface $\uent=0$, is unaffected by a smooth deformation.

One way is that if we think of Schwarzschild modes with frequencies $\omega$ and $\omega'$, then the near-horizon modes defined above pick up the coefficient of the $\delta(\omega-\omega')$ term in their two-point function. The analysis above tells us that this coefficient must be universal and cannot be changed by a smooth deformation. We will elucidate this point further in our analysis of the Reissner-Nordstr\"{o}m black hole below, where we discuss global modes and explicitly relate them to the near-horizon modes defined above.

\section{The Reissner-Nordstr\"{o}m black hole in AdS \label{secrnads}}
We now apply the considerations above to the Reissner-Nordstr\"{o}m (RN) geometry in anti-de Sitter space. The techniques that we have developed here can be just as easily applied to charged black holes in flat-space or in de Sitter space. The reason for considering AdS is not only that this allows us to make a link with the AdS/CFT conjecture but also because there is a canonical choice of a quantum state with asymptotically anti-de Sitter boundary conditions---the Hartle-Hawking state.    In flat space, it was shown  through an elaborate computation of the renormalized stress-tensor that the Hartle-Hawking state led to a singular stress-tensor on the horizon of this black hole \cite{birrell1978falling,Sela:2018xko}. Here, we will show how a similar conclusion can be reached for charged black holes in flat space much more easily using our test.

This section is divided into four parts. First, we review the features of the classical RN geometry, which also serves to introduce necessary notation. Next, we describe the global expansion of fields propagating on this background. Then we relate these global modes to the near-horizon modes described in the previous section. The criterion that the near-horizon modes should be correctly entangled leads to specific constraints on the two-point functions of the global modes. Although  we were not able to check these constraints analytically, it is easy to check them numerically in any dimension. In the last subsection, we present evidence that the Hartle-Hawking/Kruskal state is singular at the inner horizon both for asymptotically AdS RN black holes and for asymptotically flat RN black holes in various dimensions.

\subsection{The classical geometry}
The classical geometry of the RN black hole in AdS, in general spacetime dimension $d+1$, is described by the metric
\be
 ds^2 = -f(r)dt^2 +f(r)^{-1} dr^2 + r^2 d\Omega_{d-1}^2,
\ee
where 
\be
f(r) = r^2 + 1 - {A \over r^{d-2}} + {B \over r^{2(d-2)}}.
\ee
Here we have set the radius of AdS to 1 and 
and the constants $A$ and $B$ are related to the mass, $M$ and charge $Q$ by
\be
A= {16 \pi M \over (d-1) V_{d-1}}; \quad B = {(8 \pi Q)^2 \over 2 (d - 1)(d-2) V_{d-1}},
\ee
where $V_{d-1}$ is the volume of the unit $(d-1)$-sphere \cite{Chamblin:1999tk}.
We will consider non-extremal black holes where $f(r)$ has first order zeros at $\rmin$ and $\rplus$, which  are the positions of the inner and outer horizons respectively.

As usual, it is convenient to introduce the tortoise coordinate
\be
\label{tortoisedef}
dr_* = \frac{dr}{f(r)}.
\ee

Near the outer horizon we have $\rtor \rightarrow -\infty$ and we choose the origin of $\rtor$  so that as one approaches the horizon,  we have 
\be
\label{rtorrnearouter}
(r-\rplus) \rightarrow {1 \over 2 \kappa_{+}} e^{2 \kappa_{+} \rtor},
\ee
where the surface gravity at the  horizons is, as usual,  given by $\kappa_{+}=f'(r_+)/2$. This fixes $\rtor$ to a constant asymptotic value near the boundary of AdS. Moreover, near the future outer horizon, we also have $t \rightarrow \infty$ and, as usual,  spacetime can be continued past this horizon by introducing coordinates
\be
\label{uvregion1}
U = -{1 \over \kappa_{+}} e^{\kappa_{+} (r_* - t)}; \qquad V = {1 \over \kappa_{+}} e^{\kappa_{+} (r_* + t)}.
\ee
In these coordinates, the right future outer horizon is at $U = 0,V>0$. After crossing this horizon,  in the forward wedge, we define the $t,r_*$ coordinates by
\be
\label{uvregion2}
U = {1 \over \kappa_{+}} e^{\kappa_{+} (r_* - t)}; \qquad V = {1 \over \kappa_{+}} e^{\kappa_{+} (r_* + t)}.
\ee
Within the forward wedge, we can move to the left outer horizon where, again, $\rtor \rightarrow -\infty$ but $t \rightarrow -\infty$. This horizon has $V=0$. It is possible to cross this left horizon to reach a left asymptotic region.

However, in the forward wedge, it is also possible to reach the region $\rtor \rightarrow \infty$, which marks the  inner horizon. Within the forward wedge, the right inner horizon has $t \rightarrow +\infty$, and the left inner horizon has $t \rightarrow -\infty$. Near the inner horizon, the relationship between the tortoise and the ordinary radial coordinate becomes
\be
\label{rtorrinner}
r-\rmin  \rightarrow {\rconst^2 \over \kappa_{-}} e^{-2 \kappa_{-} \rtor}.
\ee
   Here, the surface gravity of the inner horizon is $\kappa_{-} = -f'(\rmin)/2$. Note that one unavoidably obtains an additional constant, $\rconst$, in the relationship between $r - \rmin$ and $\rtor$ having fixed a similar constant to $1$ near the outer horizon. Naively, it appears that a simple change of coordinates will allow the continuation of the geometry beyond the inner horizon as well, and this naive extension leads to an extended spacetime diagram. In particular, define
\be
\label{uvnearcauchy}
U' = -{\rconst \over \kappa_{-}} e^{-\kappa_{-} (\rtor + t)}  \qquad V' = -{\rconst \over \kappa_{-}} e^{\kappa_{-} (\rtor + t)}.
\ee
This places the right inner horizon at $U'=0$ and the left inner horizon at $V'=0$. When we cross the right inner horizon, we may use the coordinates
\be
U' = {\rconst \over \kappa_{-}} e^{-\kappa_{-} (\rtor + t)}  \qquad V' = -{\rconst \over \kappa_{-}} e^{\kappa_{-} (\rtor + t)}.
\ee
A similar change of coordinates allows an extension across the left inner horizon. 

\begin{figure}[!h]
\begin{center}
\includegraphics[width=0.2\textwidth]{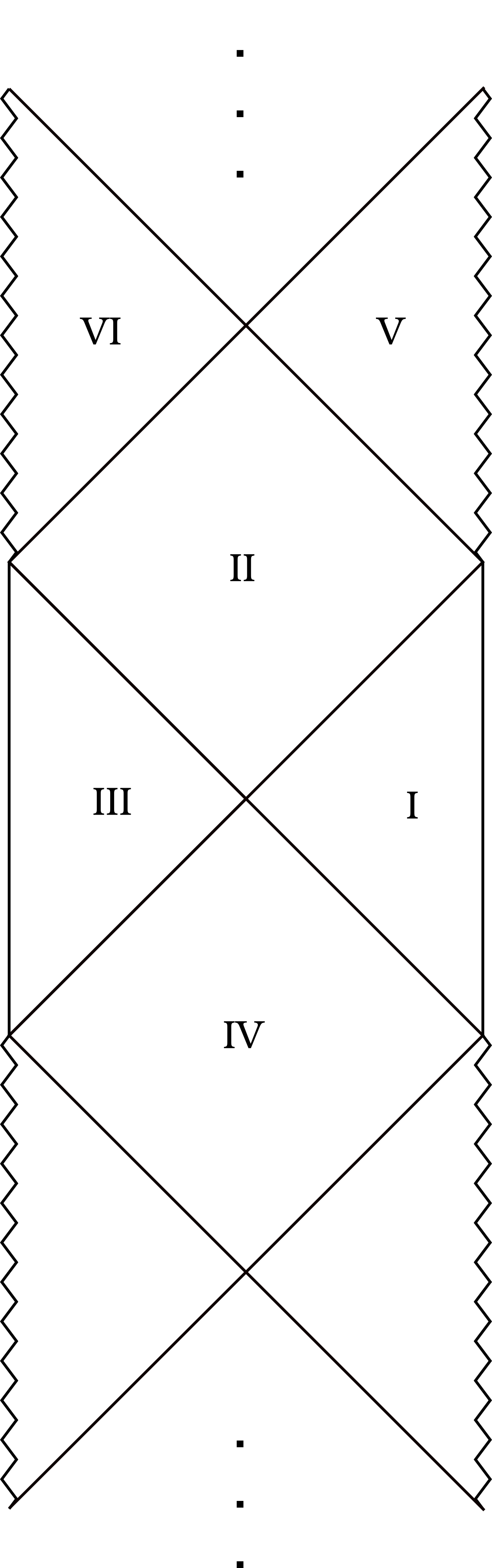}
\end{center}
\caption{\em Maximal extension of the Reissner Nordstr\"om spacetime. The dots indicate repetitions of the displayed pattern.}
\label{rnpenroseads}
\end{figure}

The naive extended Penrose diagram is shown in Figure \ref{rnpenroseads}.

\subsection{Quantum fields on the RN background}
We now consider a scalar quantum field propagating on the background above. We will assume that the scalar field is governed by an effective action
\be
\label{effectiveaction}
S_{\text{eff}} = {1\over 2}\int \sqrt{-g} \left[g^{\mu \nu}  \partial_{\mu} \phi \partial_{\nu} \phi - m^2 \phi^2 \right].
\ee
Since the field equations are linear, we expect to be able to expand the field as
\be
\label{phiexpansion}
\phi =  \sum_{\ell} \int {d \omega \over \sqrt{2 \pi}}   F_{\omega, \ell}(\rtor) e^{-i \omega t} Y_{\ell}(\Omega) + \text{h.c},
\ee
where $F_{\omega, \ell}$ are   operators. We now describe the operators $F_{\omega, \ell}$ in more detail in various limits. This also serves to define our notation.

\paragraph{Field expansion near the boundary of AdS \\}
Near the boundary of AdS, we demand that the field be  normalizable. This corresponds to a situation where no sources have been turned on for the field.  Near the boundary we have
\be
\lim_{r\rightarrow \infty} [r^{\Delta}F_{\omega, \ell}(r)] =   \op_{\omega, \ell}, 
\ee
where $\op_{\omega, \ell}$ are the modes of the boundary operator of dimension $\Delta$ that is dual to the bulk field $\phi$. 

\paragraph{Field expansion just outside the outer horizon \\}
The normalizable boundary conditions link the left and right moving modes outside the horizon. As we approach the outer horizon, the tortoise coordinate $\rtor \rightarrow -\infty$. A simple consideration of the wave-equation resulting from the action \eqref{effectiveaction} shows that as we approach the horizon from  outside, so that $r \rightarrow \rplus$ but $r > \rplus$, the field has the expansion
\be
\label{outsidehor}
F_{\omega, \ell}  \underset{r \rightarrow \rplus}{\longrightarrow} {1 \over \sqrt{2 \omega} \rplus^{d-1 \over 2}}\, a_{\omega, \ell}\, \left( e^{i \omega \rtor} + e^{-i \delta_{\omega,\ell}} e^{-i \omega \rtor} \right). 
\ee
With the normalization above, the operators $a_{\omega, \ell}$ are canonically normalized 
\be
[a_{\omega, \ell}, a_{\omega', \ell'}] = \delta(\omega - \omega') \delta_{\ell, \ell'}.
\ee

\paragraph{Field expansion just inside the outer horizon \\}
Just inside the outer horizon, the left-moving modes must be the same as the left-movers outside the horizon by continuity of the field. However, it is possible to have new right-moving modes. Therefore, as $r \rightarrow \rplus$ but $r < \rplus$, the field has the expansion
\be
\label{expansioninsideouter}
F_{\omega, \ell}  \underset{r \rightarrow \rplus}{\longrightarrow}  {1 \over \sqrt{2 \omega} \rplus^{d-1 \over 2}} \left(\ta_{\omega, \ell}^{\dagger}   e^{i \omega \rtor}  + a_{\omega, \ell}  e^{-i \delta_{\omega, \ell}}  e^{-i \omega \rtor}   \right) .
\ee

\paragraph{Field expansion just outside the inner horizon \\}
As we approach the inner horizon, we find that $\rtor \rightarrow \infty$. As we approach it from outside, i.e. $r \rightarrow \rmin$ but $r > \rmin$, we find that the radial mode functions simplify again and we may write
 \be
\label{expansionoutsideinner}
 F_{\omega, \ell} \underset{r \rightarrow \rmin}{\longrightarrow} {1 \over \sqrt{2 \omega} \rmin^{d-1\over 2}} \left(\tb_{\omega, \ell}^{\dagger}  e^{i \omega \rtor}  +  b_{\omega, \ell} \, e^{-i \omega \rtor} \right).
\ee
The modes $b$ and $\tb$ are not independent and must be related to the modes $a,\ta$ just inside the outer horizon. The relationship can be obtained by evolving the expansion given in \eqref{expansioninsideouter} forward in time until one reaches the inner horizon. We will consider this relationship in more detail in subsection \ref{secnumerics}. 

\paragraph{Field expansion just inside the inner horizon\\}
For the sake of completeness, we also describe the field expansion just inside the inner horizon. This is  an academic exercise  since we have not yet found any state where the conditions for smoothness are met even outside the inner horizon. However, if such a state were to be constructed, then the expansion behind the inner horizon would be relevant. 

Just inside the inner horizon, with $r \rightarrow \rmin$ and also $r < \rmin$, we may write
\be
F_{\omega, \ell} \underset{r \rightarrow \rmin}{\longrightarrow} {1 \over \sqrt{2 \omega} \rmin^{d-1\over 2}} \left(\tb_{\omega, \ell}^{\dagger}   e^{i \omega \rtor}  +  c^{\dagger}_{\omega, \ell}  e^{-i \omega \rtor} \right).
\ee
Here we have imposed the fact that the ``right movers'', $\tb$ cross over smoothly whereas the $c$ are some new modes which may or may not be related to the earlier modes. Notice that both $\tb$ and $c$ have the wrong energy with respect to the Schwarzschild Hamiltonian.  This is because, in our coordinates as we move up on the diagram \ref{rnpenroseads}, the value of $t$ decreases.

Note that this expansion relies only on the continuity of the metric, and does {\em not} make any assumptions about what happens deep inside the inner horizon.

\subsection{Relationship between near-horizon modes and global modes}
We can define near-horizon modes near all the horizons in this geometry. We will define modes on both sides of the right outer horizon, and on both sides of the right inner horizon. It is, of course, possible to define modes near the left-horizons but in our description (which treats the horizons symmetrically) this is redundant and so we will avoid it for now.

So we define 
\be
\label{nearinglobalpos}
\begin{split}
&\anh = {1 \over  \sqrt{\pi \omega_0}  } \int\partial_{U} \phi(U, V=0, \Omega) \left(-{U \over U_0} \right)^{-i \omega_0} \tune(-U)  Y_{\ell}^*(\Omega)\,dU d^{d-1} \Omega, \\
&\tildanh = {1 \over  \sqrt{\pi \omega_0}} \int \partial_U \phi(U,V=-\epsilon,\Omega) \left({U \over U_0} \right)^{i \omega_0} \tune(U) Y_{\ell}(\Omega) \,dU d^{d-1} \Omega, \\
&\bnh = {1 \over  \sqrt{\pi \omega_1}  } \int\partial_{U'} \phi(U', V'=0, \Omega) \left(-{U' \over U'_0} \right)^{-i \omega_1} \tune(-U') Y_{\ell}^*\, dU' d^{d-1} \Omega, \\
&\cnh = {1 \over  \sqrt{\pi \omega_1}} \int \partial_U' \phi(U',V'=-\epsilon,\Omega) \left({U' \over U'_0} \right)^{i \omega_1} \tune(U')  Y_{\ell}(\Omega) \,dU'd^{d-1} \Omega.
\end{split}
\ee
Here $\omega_1 = {\kappa_{+} \omega_0 \over \kappa_{-}}$. All these modes depend on a choice of $\omega_0$ and $\ell_0$ and while we suppress these quantities in the notation we will make some choices for them later.

The integrals in \eqref{nearinglobalpos} can be performed as follows. For the first line of \eqref{nearinglobalpos} we find that
\be
\anh= {1 \over \pi \sqrt{2 \omega_0} } \int\partial_{U} \left[F_{\omega, \ell}(\rtor) e^{-i \omega t} + F^*_{\omega, \ell}(\rtor) e^{i \omega t} \right] \left(-{U \over U_0} \right)^{-i \omega_0} \tune(-U) d U d \omega.
\ee
Now, since the tuning function has support for only very small values of $U$, we may expand the mode function using the approximation \eqref{outsidehor}. We then find that up to the $\Or[\epsilon]$ dependence on the cutoffs, described in section \ref{entanglednull},
\be
\begin{split}
\anh &= {1 \over 2 \pi \sqrt{\omega_0}} \int  a_{\omega, \ell} (\kappa_{+})^{i \omega \over \kappa_{+}}\left(\partial_{U} (-U)^{i {\omega \over \kappa_{+}}} \right) \left({-U \over U_0} \right)^{-i \omega_0 }\tune(-U) d U  {d \omega \over \sqrt{\omega}} \\
&=  {1 \over 2 \pi \sqrt{\omega_0}} \int  {i \over \kappa_{+}} \sqrt{\omega} ( \kappa_{+} U_0)^{i \omega \over \kappa_{+}} a_{\omega, \ell}  \left({-U \over U_0} \right)^{i ({\omega \over \kappa_{+}} - \omega_0)}\tune(-U) {d U \over U}  {d \omega} \\
&= i\int \sharp(\omega_0 - {\omega \over \kappa_{+}}) ( \kappa_{+} U_0)^{i  \omega \over \kappa_{+}} \sqrt{\omega \over \omega_{0}}  a_{\omega, \ell}  {d \omega \over \kappa_{+}}.
\end{split}
\ee
Note that the normalization of the modes is correct since, using the expression above, and also the identity $[a_{\omega, \ell}, a_{\omega', \ell}] = \delta(\omega - \omega')$, we can check that
\be
\begin{split}
[\anh, \anh^{\dagger}] &= \int |\sharp(\omega_0 - {\omega \over \kappa_{+}} )|^2 {\omega \over \omega_{0}} {d \omega \over \kappa_{+}^2} \\
&=\int |\sharp(\omega_0 - \omega' )|^2 {\omega' \over \omega_{0}} {d \omega'} = 1.
\end{split}
\ee
where we used that $\sharp(\omega)$ is very sharply peaked around $\omega=0$.

Similarly, we can express the near-horizon modes in terms of the global modes near the other horizons as well. We find that that in the limit under consideration precisely the same function $\sharp$ appears in these other expressions. 
\be
\label{nearinglobalfreq}
\begin{split}
&\tildanh = -i \int (\kappa_{+} U_0)^{-i  \omega \over \kappa_{+}} \sharp^*(\omega_0 - {\omega \over \kappa_{+}} ) \sqrt{\omega \over \omega_{0}}  \ta_{\omega, \ell}  {d \omega \over \kappa_{+}}. \\
&\bnh = i \int   ({\kappa_{-} \over \rconst} U_0)^{-i \omega \over \kappa_{-}} \sharp(\omega_1 - {\omega \over \kappa_{-}} ) \sqrt{\omega \over \omega_{1}}  b_{\omega, \ell}  {d \omega \over \kappa_{-}}. \\
&\cnh = -i \int  ({\kappa_{-} \over \rconst} U_0)^{i \omega \over \kappa_{-}} \sharp^*(\omega_1 - {\omega \over \kappa_{-}} ) \sqrt{\omega \over \omega_{1}}  c_{\omega, \ell}  {d \omega \over \kappa_{-}}. \\
\end{split}
\ee
As advertised, in each case the near-horizon modes can be written as global modes smeared with a function that is  sharply peaked in frequency space.

\subsection{Constraints on two-point functions}

From the results of section \ref{entanglednull}, we now find the following constraints on the near-horizon modes defined above. As explained there, smoothness of the outer horizon implies that
\be
\begin{split}
&\langle \Psi | \tildanh \tildanh^{\dagger} | \Psi \rangle = \langle \Psi | \anh \anh^{\dagger} | \Psi \rangle = {1 \over 1 - e^{-2 \pi \omega_0}}; \\
&\langle \Psi | \tildanh \anh | \Psi \rangle = \langle \Psi | \tildanh^{\dagger} \anh^{\dagger} | \Psi \rangle = {e^{-\pi \omega_0} \over 1 - e^{-2 \pi \omega_0}}. 
\end{split}
\label{nearohoccupation}
\ee
On the other hand, upon applying these constraints near the inner horizon we find that
\be
\begin{split}
&\langle \Psi | \bnh \bnh^{\dagger} | \Psi \rangle = \langle \Psi | \cnh \cnh^{\dagger} | \Psi \rangle = {1  \over 1 - e^{-2 \pi \omega_1}}; \\
&\langle \Psi | \bnh \cnh | \Psi \rangle = \langle \Psi | \bnh^{\dagger} \cnh^{\dagger} | \Psi \rangle = {e^{-\pi \omega_1}  \over 1 - e^{-2 \pi \omega_1}},
\end{split}
\label{nearihoccupation}
\ee
where $\omega_1 = {\kappa_{+} \over \kappa_{-}} \omega_0$ as above. 

Now, as discussed earlier, the modes $\bnh,\bnh^\dagger$ can in principle be expressed as linear combinations of $\anh,\tildanh,\anh^\dagger, \tildanh^\dagger$ by solving the wave equation in the region between the two horizons.  Hence it is not obvious whether equations \eqref{nearohoccupation} and \eqref{nearihoccupation} can hold simultaneously. In fact as we will show numerically in the next subsections, these equations are incompatible in the case of the Hartle-Hawking state for the AdS-RN black hole. This statement implies that this state is singular on the inner horizon of the AdS-RN black hole.

Before we continue with this analysis, let us see what the constraints \eqref{nearohoccupation}, \eqref{nearihoccupation} imply for the  two-point function of the global modes $a, b,\ta$. Let us assume that the two-point function of the global modes is given by 
\be
\begin{split}
&\langle \Psi | a_{\omega, \ell} a^{\dagger}_{\omega', \ell} | \Psi \rangle =  {\cal A}_1(\omega) \delta(\omega - \omega') + {\cal A}_2(\omega, \omega'); \\
&\langle \Psi | \ta_{\omega, \ell} \ta^{\dagger}_{\omega', \ell} | \Psi \rangle =  \widetilde{{\cal A}}_1(\omega) \delta(\omega - \omega') + \widetilde{{\cal A}}_2(\omega, \omega') ;\\
&\langle \Psi | a_{\omega, \ell} \ta_{\omega', \ell} | \Psi \rangle =  {\cal C}_1(\omega) \delta(\omega - \omega') + {\cal C}_2(\omega, \omega'); \\
&\langle \Psi | b_{\omega, \ell} b^{\dagger}_{\omega', \ell} | \Psi \rangle =  {\cal B}_1(\omega) \delta(\omega - \omega') + {\cal B}_2(\omega, \omega'),
\end{split}
\ee
where, in each case, we have separated the correlator into one part that is proportional to a delta function and another part that is assumed to be smooth at $\omega = \omega'$. (We have also used the spherical symmetry to set the same value for $\ell$ in all the operators above.) Then, substituting the formulas above, we find that the coefficients of the delta function are  completely fixed by demanding smoothness at the outer and the inner horizon. In particular, we find that
\be
\label{globalconstraints}
\begin{split}
&{\cal A}_1(\omega) = \widetilde{{\cal A}}_1(\omega) = {1 \over 1 - e^{-{2 \pi \over \kappa_{+}} \omega}} ; \\
&{\cal C}_1(\omega) = {e^{-{\pi \over \kappa_+} \omega}  \over 1 - e^{-{2 \pi \over \kappa_{+}} \omega}};   \\
&{\cal B}_1(\omega) =  {1 \over 1 - e^{-{2 \pi \over \kappa_{-}} \omega}},
\end{split}
\ee
where we have used the relationship between $\omega_1 = \kappa_{+} \omega_0/\kappa_{-}$. Note that the factors that appear in the exponentials above are just the standard inverse-temperatures of the inner and outer horizons given by $\beta_{\pm} = {2 \pi \over \kappa_{\pm}}$.

Therefore we see that the two-point function of the near-horizon modes picks only the delta-function piece in the two-point function of the global modes and completely fixes that piece.  This explains why a smooth deformation of the geometry cannot change the two-point function of the near-horizon modes: a smooth deformation of the geometry may change the smooth part of the two-point function of the global modes, but it cannot alter the coefficient of the delta function in this two-point function. This is the only term that the near-horizon modes are sensitive to and it must take a given value near a smooth horizon.

Now, as we explained above, in any given state of the system, the $b$ modes are obtained by evolving the $a$ and $\ta$ modes in the region between the inner and outer horizon of the black hole. Therefore, given the correlators of those modes, it is possible to check whether the constraints \eqref{globalconstraints} are satisfied. We will show that for the RN geometry, in various dimensions, these constraints cannot all be satisfied simultaneously.

\subsection{Numerical results for the Reissner-Nordstr\"{o}m black hole \label{secnumerics}}
We now check whether our constraints  are satisfied in the AdS Reissner-Nordstr\"{o}m geometry. 

In this situation, where the propagation is linear, and using the time-translation isometry of the geometry we can write
\be
\bnh = \boga  \anh + \bogta \tildanh^{\dagger}.
\ee
It is the time-translation invariance that allows us to relate modes with the same value of $\omega$, and we note that this equation may be modified if interactions are strong, or if the geometry is strongly time-dependent. 
Therefore,
\be
\label{bnhocc}
\begin{split}
\langle \bnh \bnh^{\dagger} \rangle &= \langle \anh \anh^{\dagger} \rangle |\boga|^2 + |\bogta|^2 \langle \tildanh^{\dagger} \tildanh \rangle + 2 \text{Re} \left(\boga \bogta^{*} \langle \anh \tildanh \rangle \right) \\
&= {1 \over 1 - e^{-2 \pi \omega_0}} \left(|\boga|^2 + e^{-2 \pi \omega_0} |\bogta|^2  + 2 e^{-\pi \omega_0} \text{Re}\left(\boga \bogta^* \right) \right). 
\end{split}
\ee
So the question of checking whether the near-horizon constraints  \eqref{nearohoccupation},\eqref{nearihoccupation} are satisfied as we approach the near-horizon reduces to evaluating the Bogoliubov coefficients, $\boga, \bogta$ and determining if the combination above is in agreement with \eqref{nearihoccupation}.By computing the two-point function of the modes near the inner horizon we define
\be
\delta = {\langle \bnh^{\dagger} \bnh \rangle (1 - e^{-2 \pi \omega_1})} - 1.
\ee
which is the  fractional difference from the expected Boltzmann factor.  If the constraints are satisfied we will have $\delta = 0$.

\paragraph{Description of the numerical algorithm \\}
These Bogoliubov coefficients can be computed by solving the radial  (ordinary) differential equation that fixes the evolution of the field. We briefly describe our algorithm. A solution of the wave-equation with frequency $\omega$ and angular momentum, $\ell$ can be written in the form $\phi_{\omega, \ell}(\rtor) e^{-i \omega t} Y_{\ell}(\Omega)$. The radial part of the solution $\phi_{\omega, \ell}$ obeys the equation
\be
\label{waveeqn}
{d \over d \rtor} \left(r^{d - 1} {d \phi_{\omega, \ell} \over d \rtor}\right) + \omega^2 \phi_{\omega, \ell} - 
 \ell (\ell + d - 2) f(r) {\phi_{\omega, \ell} \over r^2} - m^2 \phi_{\omega, \ell} f(r) = 0,
\ee
where $r$ is determined by solving the auxiliary equation $r'(\rtor)=f(r)$. This auxiliary equation must be solved first; demanding the behaviour \eqref{rtorrnearouter} near the outer horizon allows us to fix the value of $\rtor$ near the boundary of AdS as $r \rightarrow \infty$. Solving the auxiliary equation between the inner and outer horizon allows us to determine $\rconst$, which appears in \eqref{rtorrinner}.

The equation \eqref{waveeqn} is now solved, as a set of two first-order differential equations, starting from the boundary with initial conditions set by $\phi_{\omega, \ell}(r) \rightarrow {1 \over r^{\Delta}}$ at large $r$. As $r \rightarrow \rplus$, we can then read off the phase $\delta_{\omega, \ell}$ by matching the behaviour of $\phi_{\omega, \ell}$ and $\phi'_{\omega, \ell} = {d \phi_{\omega, \ell} \over d \rtor}$  to \eqref{outsidehor}.

Now, starting at a point in the middle of the inner and outer horizons, we solve the equation towards $\rtor \rightarrow -\infty$ (the outer horizon) and also $\rtor \rightarrow \infty$ (the inner horizon) with two separate initial conditions: once with $\phi_{\omega, \ell} = 0, \phi'_{\omega, \ell} = 1$ and a second time with $\phi_{\omega, \ell} = 1, \phi'_{\omega, \ell} = 0$. Near both the inner and the outer horizons, the asymptotic behaviour of the mode is given by \eqref{expansioninsideouter} and \eqref{expansionoutsideinner}. By determining the coefficients of the terms that oscillate as $e^{\pm i \omega \rtor}$ for both sets of initial conditions, we can solve for the Bogoliubov coefficients $\boga$ and $\bogta$. This immediately leads to a result for $\delta$.

The procedure above is quite simple, and can be implemented using any standard numerical library.   In our own case, to solve the differential equations, we used an eighth-order explicit Runge-Kutta solver with the Dormand and Prince coefficients \cite{hairer1993sod} as implemented in C using code from CALCODE \cite{calcode1999}. We also used the root-finding routines in the GNU scientific library \cite{galassi2009gnu} to locate the horizons and GNU parallel \cite{Tange2011a} to parallelize the calculations.

\begin{figure}[!h]
\begin{center}
\begin{subfigure}{0.4\textwidth}
\includegraphics[width=\textwidth]{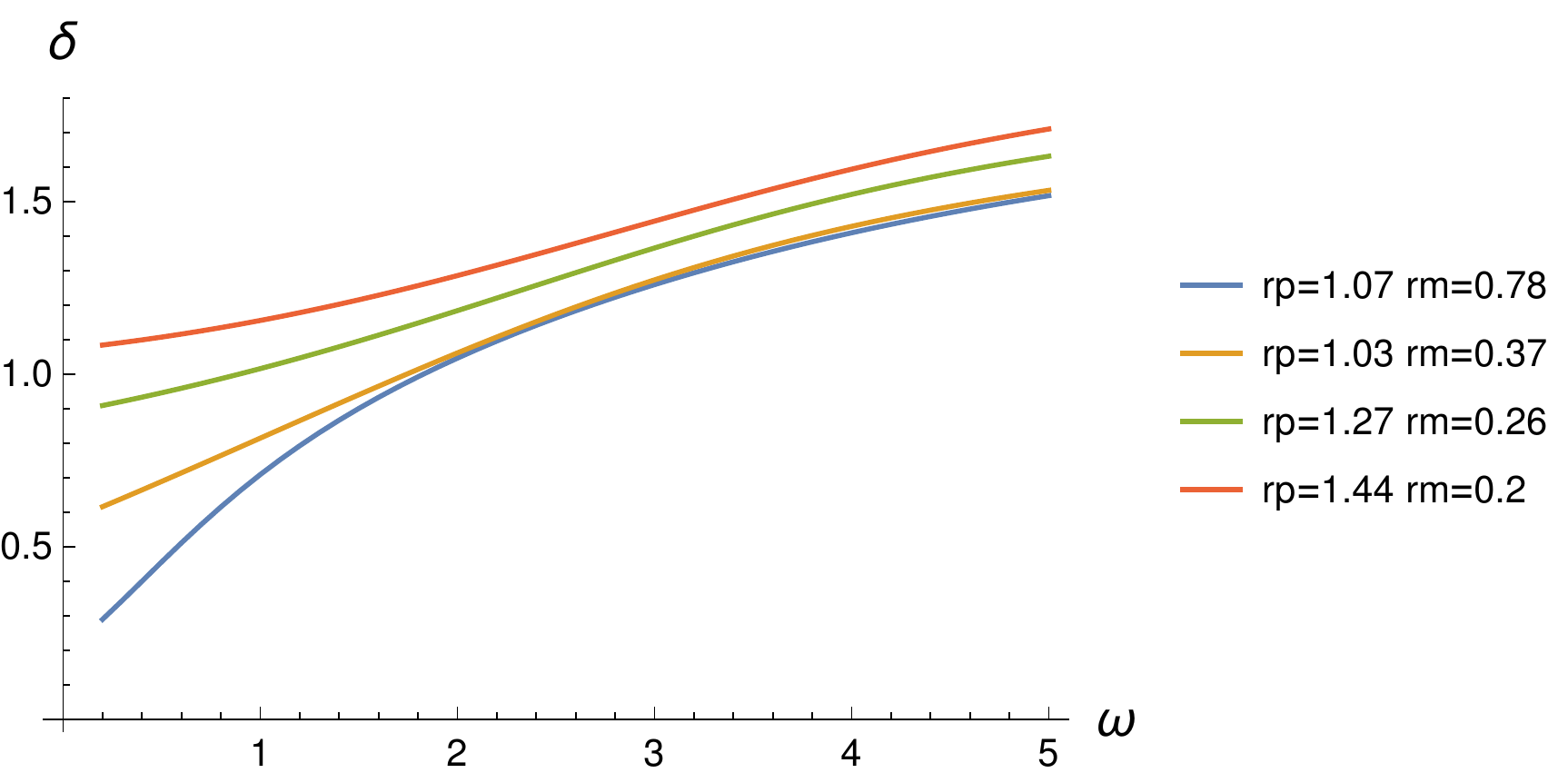}
\caption{$d=3$}
\end{subfigure}
\begin{subfigure}{0.4\textwidth}
\includegraphics[width=\textwidth]{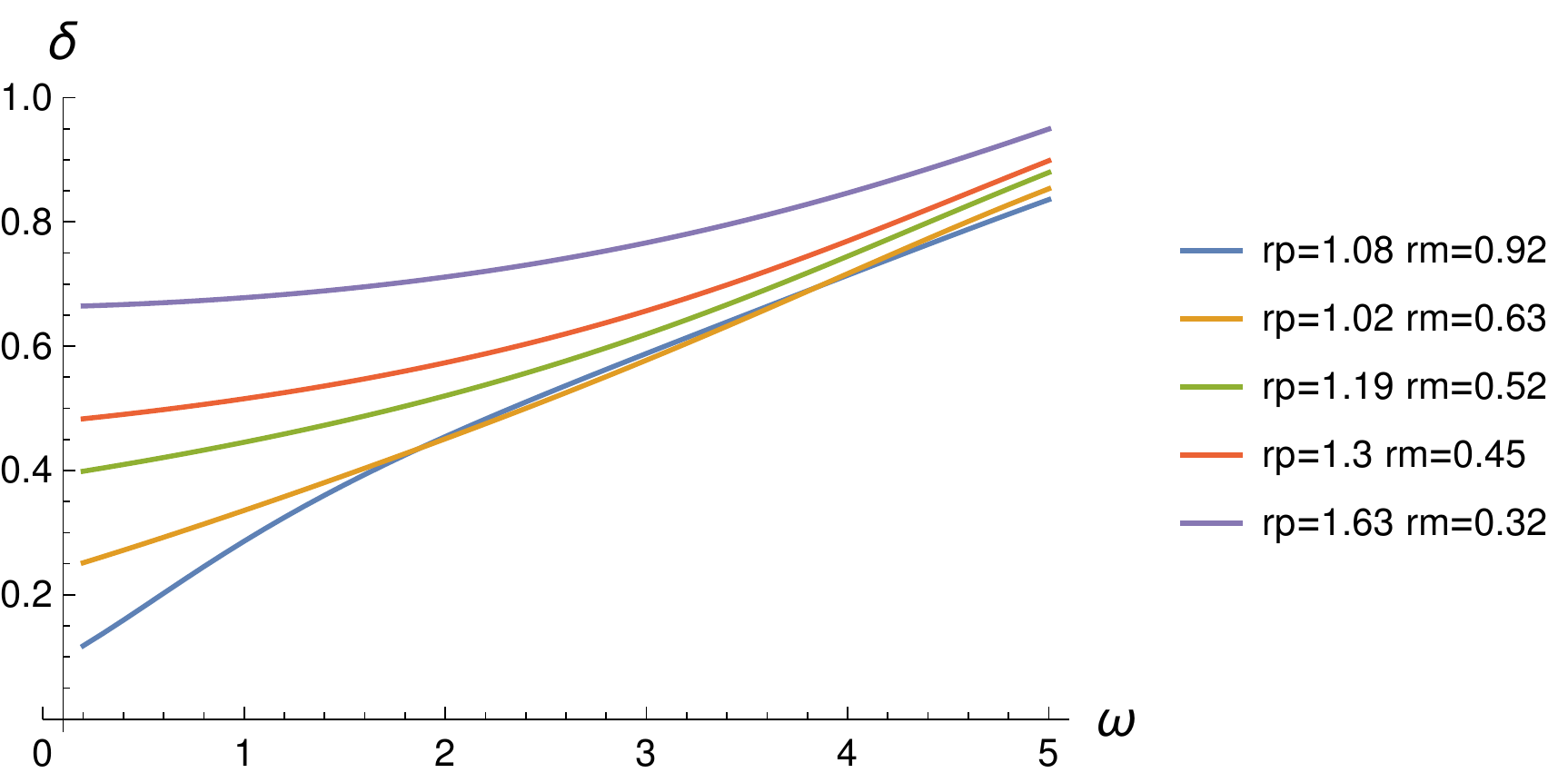}
\caption{$d=4$}
\end{subfigure}
\begin{subfigure}{0.4\textwidth}
\includegraphics[width=\textwidth]{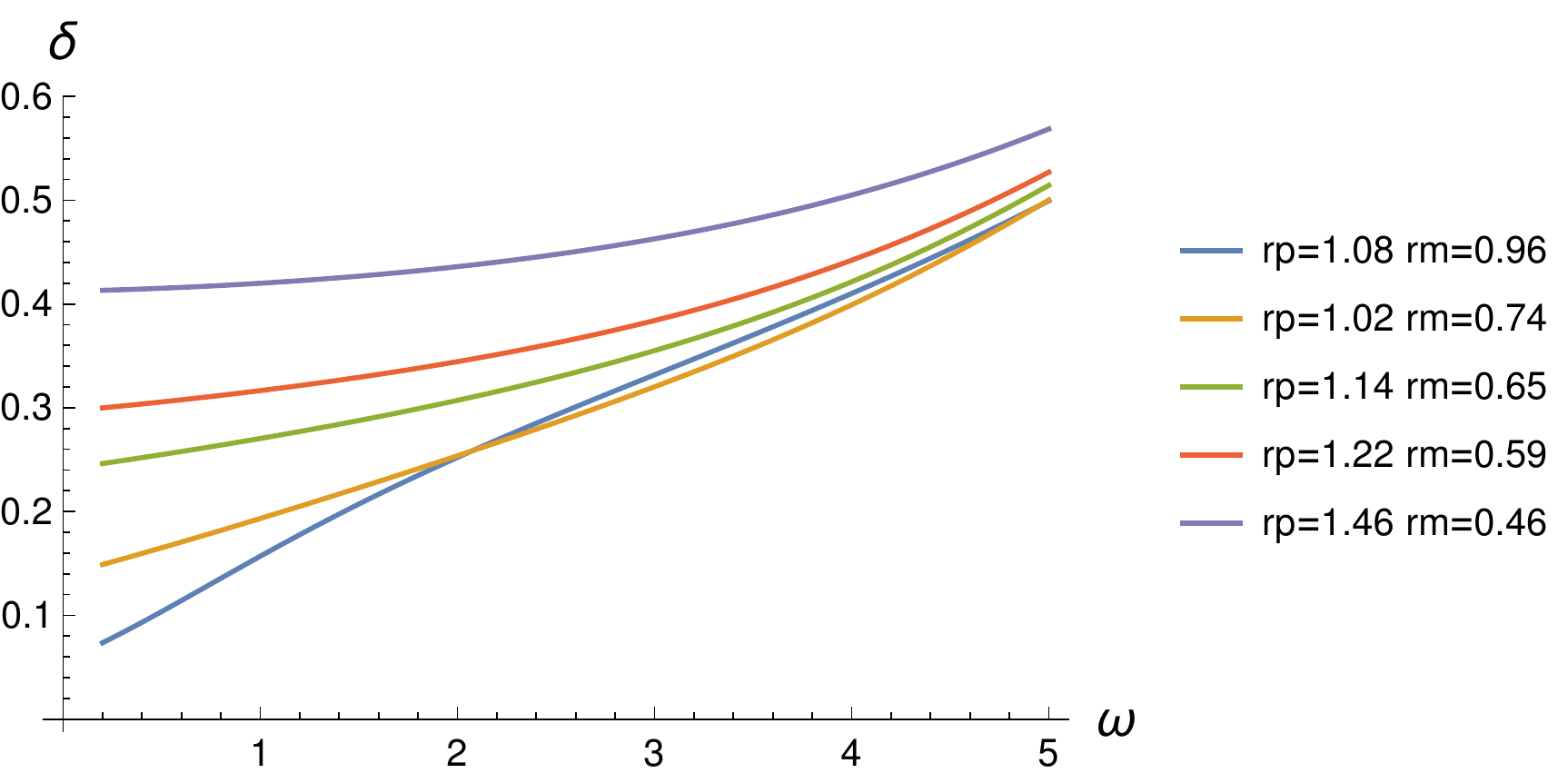}
\caption{$d=5$}
\end{subfigure}
\begin{subfigure}{0.4\textwidth}
\includegraphics[width=\textwidth]{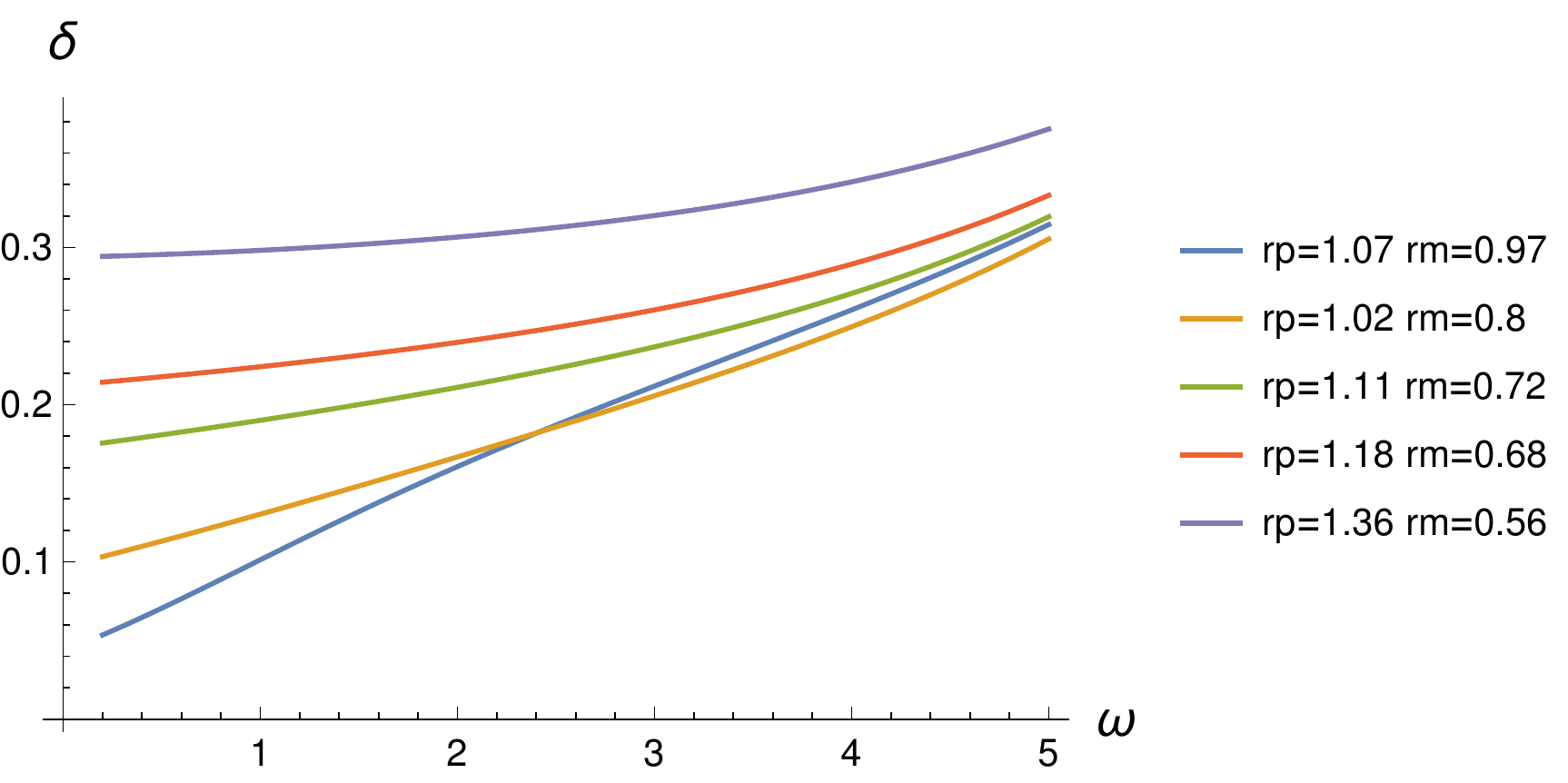}
\caption{$d=6$}
\end{subfigure}
\caption{\em A plot of the fractional difference, $\delta$, vs the frequency for $\ell=0$ for various values of the inner and outer temperature. Non-zero values of $\delta$ indicate that the constraints of section \ref{entanglednull} are not satisfied. \label{fracdiffvswplot}}
\end{center}
\end{figure}

\paragraph{Results \\}
In Figure \ref{fracdiffvswplot} we present a plot for $\delta$ vs the frequency for the angular momentum $\ell = 0$ for various values of the radii of the inner and outer horizon and in dimensions $d = 3,4,5,6$. The black holes range from those that are near-extremal to those where there is a significant separation of scales between the two horizons.  The AdS radius is fixed to unity. 

These graphs are already sufficient to show that the Hartle-Hawking state is singular at the inner horizon of these geometries. This is because the criterion of section \ref{entanglednull} must be satisfied for each value of frequency and angular momentum, as a necessary condition, for the state to be smooth. Instead the graphs above all display non-zero values of $\delta$ for generic choices of frequency and angular momentum.

Our numerical results are quite robust. In fact, the largest source of numerical error arises from the relative phase between $\boga$ and $\bogta$ in \eqref{bnhocc}. This phase requires a careful determination of $\rconst$ and a treatment of the wave-equation near the horizon. However, even this is not a significant source of error since, unlike the case of \cite{Lanir:2018vgb},  we are considering a specific mode and so the wave-equation simplifies greatly near the horizon. Moreover, it is not difficult to check that if one keeps the angular momentum fixed and increases $\omega$ then beyond a point, {\em no choice of relative phase} between $\boga$ and $\bogta$ will yield $\delta = 0$. This allows us to reach the physical conclusions above --- that the inner horizon is not smooth in the Hartle-Hawking state --- with confidence.

\begin{figure}[!h]
\begin{center}
\includegraphics[width=0.6\textwidth]{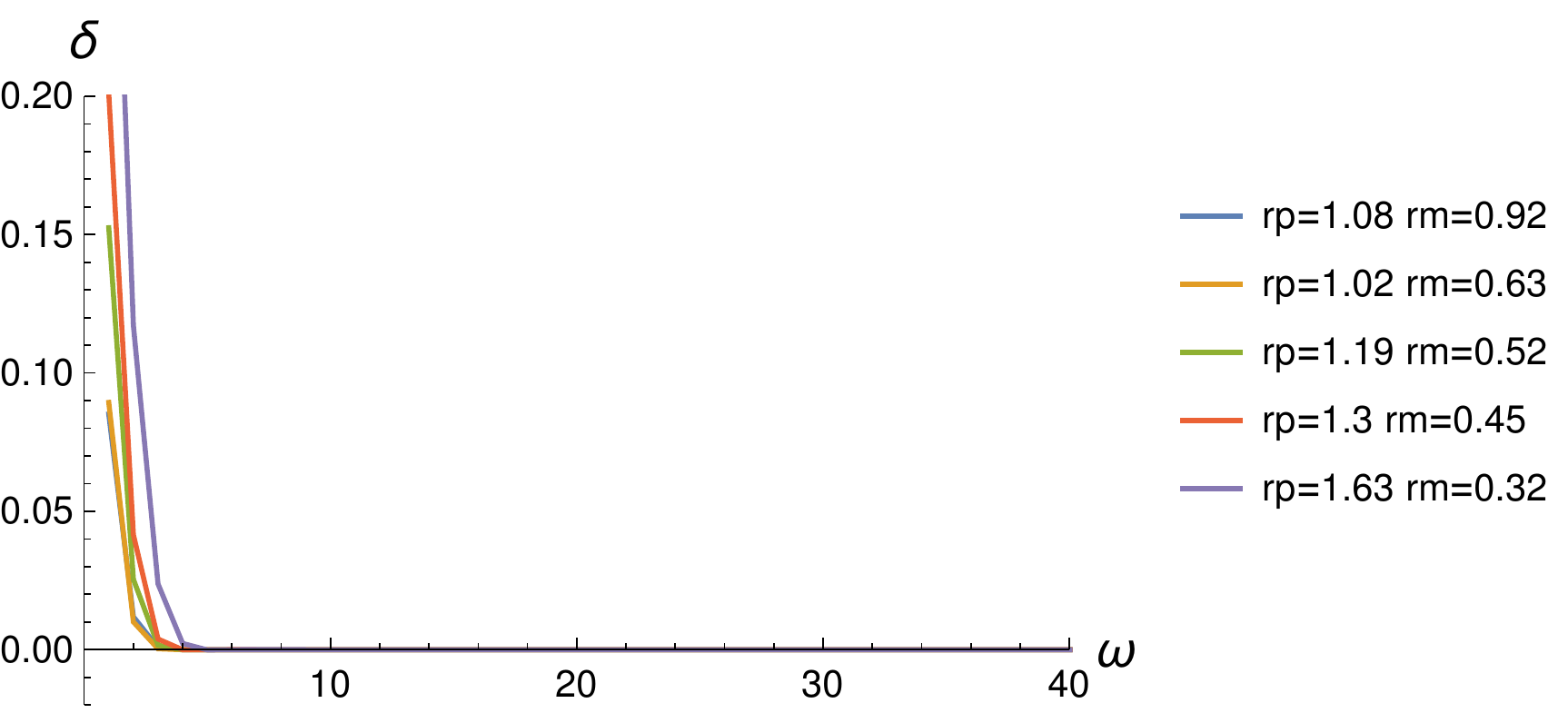}
\caption{\em A plot of the fractional difference, $\delta$, vs the angular momentum with frequency fixed to the AdS scale  for various sizes of the inner and outer horizons. \label{fracdiffvslplot}}
\end{center}
\end{figure}

However, it is also interesting to understand how the fractional difference varies with angular momentum. In Figure \ref{fracdiffvslplot}, we present a plot for $\delta$ vs the angular momentum for frequency fixed at the AdS scale for $d=4$. The reader will note the remarkable fact that at large $\ell$, the fractional difference tends to zero. This can be explained via a semiclassical WKB analysis in this limit, as we demonstrate in Appendix \ref{applargel}. This WKB analysis serves as an additional check on our numerical algorithm.

We note that  in \cite{Sela:2018xko}, it was pointed out that the renormalized expectation value of  $\phi^2$ is less singular than expected as one approaches the inner horizon. This is directly a consequence of the large $\ell$ behaviour of Figure \ref{fracdiffvslplot}, since the fastest divergence of $\phi^2$ as one approaches the inner horizon is controlled entirely by the large-$\ell$ modes.

\section{The BTZ black hole \label{secbtz}}

We now turn to a discussion of the rotating BTZ black hole. It was argued in \cite{Dias:2019ery} that a rotating BTZ black hole that is close enough to extremality violates strong cosmic censorship, and so this presents an excellent test-case for our criterion.

We start by reviewing the geometry of the BTZ black hole, and the propagation of quantum fields in this geometry. We then show that it is indeed the case that the near-horizon modes, as one approaches the inner horizon from the outside, have the correct occupation number. This corresponds to the fact that if we set the local temperature of the field correctly near the outer horizon,  it is automatically red-shifted by the geometry so that the local temperature of the modes near the inner horizon coincides exactly with the temperature of the inner horizon! This is in contrast to what we found numerically in the previous section for the AdS-RN black hole in higher dimensions.

However, as explained in section \ref{entanglednull}, if we want to extend the spacetime behind the inner horizon, we also require the existence of modes  behind the inner horizon. The reader may worry that the extension of quantum fields behind the inner horizon is not unique. Nevertheless, as we have emphasized  {\em the near-horizon modes just behind the inner horizon are fixed by requirements of smoothness} and do not require knowledge of the dynamics deep behind the inner horizon.

When one crosses the outer horizon, interior modes  can be understood using the standard construction of the mirror operators \cite{Papadodimas:2013jku}. However, we show that this construction  fails at the inner horizon due to the monogamy of entanglement. In particular, since the modes between the inner and the outer horizon are already entangled with the modes outside the outer horizon, they cannot also be entangled with new modes behind the inner horizon. 

Nevertheless we show, remarkably, that it is possible to reuse the modes between the inner horizon and the outer horizon behind the inner horizon as modes for the field behind the inner horizon. This is dependent on the fact that the modes outside the inner horizon are correctly populated. The modes that we write down uniquely fix the behaviour of the field just behind the inner horizon although the usual ambiguities associated with a Cauchy horizon arise if we attempt to probe deeper into the geometry.

We will follow the notation of \cite{Dias:2019ery} in large part. 
 
\subsection{Classical geometry and propagation of fields}
The BTZ solution can be written in the form
\be
ds^2 = -f(r) dt^2 + {d r^2 \over f(r)} + r^2 (d\phi - \Omega(r) dt)^2,
\ee
where 
\be
f(r) = {(r^2 - \rplus^2) (r^2 - \rmin^2)  \over r^2}; \qquad \Omega(r) = {\rplus \rmin \over r^2}.
\ee
Here we have set the AdS radius to 1. 

The positions of the inner and outer horizon are at $\rplus$ and $\rmin$. The angular velocities $\Omega_{\pm}$ and surface gravities $\kappa_{\pm}$ of the two horizons are given by 
\be
\Omega_{\pm}={r_{\mp} \over r_{\pm}};\qquad \kappa_{\pm} = {r_+^2 -r_-^2 \over r_{\pm}}.
\ee

The tortoise coordinate is defined as in \eqref{tortoisedef} and we once again adopt the convention that as one approaches the outer horizon, $(r - \rplus) = {1 \over \kappa{+}} e^{2 \kappa_{+} \rtor}$. We can check that near the inner horizon this leads to a relationship of the form \eqref{rtorrinner} with
\be
\rconst = 2^{\frac{\rmin+\rplus}{2 \rmin}} (\rplus-\rmin)^{\frac{\rplus}{\rmin}+1}.
\ee
We note that our conventions for $r_*$ differ from those of \cite{Dias:2019ery} by an overall additive constant.

We now consider a minimally coupled scalar field $\phi$ propagating in this geometry. Just as above, we may expand the field  in various regions. However, the expansion in terms of creation and annihilation operators is slightly more
subtle than in the non-rotating case, as we explain below. 

Near the boundary of AdS we may, as usual, expand the field as
\be
\phi \underset{r \rightarrow \infty}{\longrightarrow}{1\over 2\pi} \sum_m \int {d \omega}  {{\cal O}_{\omega, m} \over r^{\Delta}} e^{-i \omega t} e^{i m \phi} + \text{h.c},
\label{normalizability}
\ee
where 
 $\Delta (\Delta - 2) = m^2$ and ${\cal O}_{\omega, m}$ are the Fourier modes of the primary operator of dimension $\Delta$ that is dual to the field $\phi$.

Near the outer horizon, as we approach it from outside on the right, the expansion of the field $\phi$ is\footnote{We find it more convenient notationally to define the modes $a_{\omega_+,m}$ in a somewhat asymmetric conventions, so that there is no phase factor $e^{i\delta_{\omega_+,m}}$ in front of $e^{i \omega_+ \kappa_{+} r_*}$.}
\be
\label{nearouter}
\phi \underset{r \rightarrow \rplus}{\longrightarrow} \sum_m \int {d \wp  \over 2 \pi \sqrt{\rplus} \sqrt{2 \wp}} a_{\wp,m} \left(e^{i \wp \kappa_{+} \rtor}  + e^{-i \delta_{\wp, m}} e^{-i \wp \kappa_{+} \rtor} \right) e^{-i \kappa_{+} \wp t} e^{i m (\phi - \Omega_{+} t)} + \text{h.c},
\ee
where
\be
\wp = {\omega - m \Omega_{+} \over \kappa_{+}} ; \qquad \wm = {\omega - m \Omega_{-} \over \kappa_{-}}.
\label{defomegaplus}
\ee
A subtlety here is that in the expression above, the question of whether the coefficient of the mode function is a creation or annihilation operator is determined by the positivity of $\wp$ and {\em not} of $\omega$. With this convention the operators above are canonically normalized
\be
[a_{\wp,m}, a_{\wp',m'}^{\dagger}] = \delta(\wp - \wp')\delta_{mm'}.
\ee

The expansions \eqref{normalizability} and \eqref{nearouter}   fix the relationship between the operators $a_{\omega_+,m}$ and ${\cal O}_{\omega,m}$ and also the phase $\delta$. These can be determined by solving the wave-equation between the boundary and the outer horizon, which can be done analytically \cite{Balasubramanian:2004zu,KeskiVakkuri:1998nw}. We will use the solutions as written in the conventions of \cite{Dias:2019ery}.
\be
\label{Cdeltaformula}
\begin{split}
&a_{\wp, m} = {1 \over C} {\cal O}_{\omega, m} \\
&C = \left({\kappa_{+} \over \sqrt{2}} \right)^{i \omega_{+}} (\rplus^2 - \rmin^2)^{\Delta \over 2}{1 \over \sqrt{r_+} \sqrt{2 \omega_+} }{\Gamma\left({1\over 2}(\Delta - i \omega_+ - i \omega_-)\right)\Gamma\left({1\over 2}(\Delta - i \omega_+ + i \omega_-)\right) \over \Gamma(\Delta)\Gamma(-i\omega_+)}
\\
   &e^{-i \delta} = \left({\kappa_{+} \over \sqrt{2}} \right)^{2 i \omega_{+}} \frac{\Gamma (i \omega_+) \Gamma \left(\frac{1}{2}  (\Delta - i \omega_{+} - i \omega_-)\right) \Gamma \left(\frac{1}{2} (\Delta - i \omega_{+} + i \omega_- )\right)}{\Gamma (-i \omega_+) \Gamma \left(\frac{1}{2} 
   (\Delta + i\omega_{+}- i \omega_- )\right) \Gamma \left(\frac{1}{2} (\Delta  + i \omega_{+}+i \omega_-)\right)}.
\end{split}
\ee
We now proceed with the expansion of the field just inside the outer horizon
\be
\label{insideouter}
\begin{split}
\phi \underset{r \rightarrow \rplus}{\longrightarrow} &\sum_m \int {d \wp \over 2 \pi \sqrt{\rplus} \sqrt{2 \wp}} \left(a_{\wp,m} e^{-i \kappa_{+} \wp t} e^{i m (\phi - \Omega_{+} t)} + \ta_{\wp,m} e^{i \kappa_{+} \wp t} e^{-i m (\phi - \Omega_{+} t)} \right) e^{-i \wp \kappa_{+} \rtor}  \\
&+ \text{h.c}.
\end{split}
\ee
Once we are near the inner horizon, we may write
\be
\label{outsideinner}
\begin{split}
\phi \underset{r \rightarrow \rmin }{\longrightarrow} &\sum_m \int {d \wm \over 2 \pi \sqrt{\rmin} \sqrt{2 \wm}} \left(b_{\wm,m} e^{-i \kappa_{-} \wm t} e^{i m (\phi - \Omega_{-} t)} + \tb_{\wm,m} e^{i \kappa_{-} \wm t} e^{-i m (\phi - \Omega_{-} t)} \right) e^{-i \wm \kappa_{-} \rtor}  \\
&+ \text{h.c},
\end{split}
\ee
where
\be
\wm ={ \omega - m \Omega_{-} \over \kappa_{-}}.
\ee
Note that in the expansion near the inner horizon, the classification of operators into creation and annihilation operators is determined by the sign of $\wm$. 
Finally, if the field extends across the inner horizon, we may write
\be
\label{insideinner}
\phi \underset{r \rightarrow \rmin}{\longrightarrow} \sum_m \int {d \wm \over 2 \pi \sqrt{\rmin} \sqrt{2 \wm}} \left( c_{\wm,m} e^{i \wm \kappa_{-} \rtor}   + \tb_{\wm,m} e^{-i \wm \kappa_{-} \rtor}  \right) e^{i \kappa_{-} \wm t} e^{-i m (\phi - \Omega_{-} t)}    + \text{h.c}.
\ee

\subsection{Near-horizon modes and constraints from entanglement}
We now describe the relationship between the near-horizon modes and the global modes described above, and explain the constraints that the analysis of section \ref{entanglednull} places on the two-point functions of the global modes. 

Here, it is more convenient to use Eddington-Finkelstein coordinates, rather than Kruskal coordinates. These coordinates can be defined in the vicinity of both the inner and the outer horizon. Near the outer horizon, we set $v_{+}=\rtor + t$ so that the metric is given by 
\be
ds^2 = -f(r) d v_{+}^2 + 2 dv_{+} dr + r^2(d \phi  - \Omega(r) d t)^2.
\ee 
Now define $x_{+} = \rplus - r$ and $\theta_{+} = \phi - \Omega_{+}  t.$ Then, as we approach the horizon which is at $x_{+} = 0$,  the metric becomes 
\be
ds^2 = -2 dv_+ d x_{+} + \rplus^2 d \theta_{+}^2 + \Or[x_{+}] ,
\ee
which is the form that was required in section \ref{entanglednull}. 

We now define near-horizon modes using the general prescription outlined in \ref{entanglednull}. In particular, define 
\be
\begin{split}
&\anh = {1 \over  \sqrt{\pi \omega_0} } \int\partial_{x_{+}} \phi(x_{+}, v_{+}=0, \theta_{+}) \left(-{x_{+} \over U_0} \right)^{-i \omega_0} \tune(-x_{+}) d x_{+}  e^{-i m \theta_{+}} {d \theta_{+}} \sqrt{r_{+} \over 2 \pi}, \\ 
&\tildanh = {1 \over  \sqrt{\pi \omega_0} } \int\partial_{x_{+}} \phi(x_{+}, v_{+}=-\epsilon, \theta_{+}) \left(-{x_{+} \over U_0} \right)^{i \omega_0} \tune(x_{+}) d x_{+} e^{i m \theta_{+}} d \theta_{+} \sqrt{r_{+} \over 2 \pi}.
\end{split}
\ee
Such modes can be defined for any value of $v_{+}$ but, as the reader can see from the relationship with the global modes below, shifting the value of $v_{+}$ only rescales the mode by a phase.

Similarly, we can define near-horizon modes near the inner horizon. There, with $v_{-}=\rtor-t$, and $x_{-}=\rmin-r$, and  $\theta_{-} = \phi - \Omega_{-}  t$, as we approach the inner horizon at $r_{-}=0$, the metric becomes
\be
ds^2 = -2 dv_- d x_{-} + \rplus^2 d \theta_{-}^2 + \Or[x_{-}].
\ee
Therefore, the prescription of section \ref{entanglednull} tells us that near-horizon modes can be defined near the inner horizon using
\be
\begin{split}
&\bnh = {1 \over  \sqrt{\pi \omega_1} } \int\partial_{x_{-}} \phi(x_{-}, v_{-}=0, \theta_{-}) \left(-{x_{-} \over U_0} \right)^{-i \omega_1} \tune(-x_{-}) d x_{-}  e^{i m \theta_{-}} d \theta_{-} \sqrt{r_{-} \over 2 \pi}, \\ 
&\cnh = {1 \over  \sqrt{\pi \omega_1} } \int\partial_{x_{-}} \phi(x_{-}, v_{-}=-\epsilon, \theta_{-}) \left(-{x_{-} \over U_0} \right)^{i \omega_1} \tune(x_{-}) d x_{-}  e^{-i m \theta_{-}} d \theta_{-} \sqrt{r_{-} \over 2 \pi}.
\end{split}
\ee
As in the section above,  $\omega_1 = {\kappa_{+} \omega_0 \over \kappa_{-}}$.

Using the expansion of the field near the various horizons, we can relate these modes to the global modes. In doing the relevant integrals, the reader should keep in mind that the expressions above, which are given in terms of $\rtor$ and $t$ need to be transformed to $x_{\pm}, v_{\pm}$. So, for instance, 
\be
e^{i \kappa_{+} \omega_{+} (\rtor - t)} = e^{2 i \kappa_{+} \omega_{+} \rtor} e^{-i \kappa_{+} \omega_{+} v_{+}} = (\kappa_{+} x_{+})^{i \omega_{+}} e^{-i \kappa_{+} \omega_{+} v_{+}}.
\ee
Making similar substitutions in the other near-horizon expansions and using the definition \eqref{sharpdef}, we find
\be
\begin{split}
&\anh= i \int \sharp(\omega_{0} - \omega_{+} ) a_{\omega_{+},m}  (\kappa_{+} U_0)^{i \omega_{+}}  \sqrt{\omega_{+} \over \omega_{0}}  d \omega_{+};  \\
&\tildanh= -i \int \sharp^*(\omega_{0} - \omega_{+}) \ta_{\omega_{+},m}   (\kappa_{+} U_0)^{-i \omega_{+}}  \sqrt{\omega_{+} \over \omega_{0}}  d \omega_{+};  \\
&\bnh= i \int \sharp(\omega_1 - \omega_{-}) b_{\omega_{-},m}  ({\kappa_{-} U_0 \over \rconst})^{i \omega_{-}}  \sqrt{\omega_{-} \over \omega_{0}}  d \omega_{-} ;\\
&\cnh= -i \int \sharp^*(\omega_1 - \omega_{-} ) c_{\omega_{-},m}   ({\kappa_{-} U_0 \over \rconst})^{-i \omega_{-}}  \sqrt{\omega_{-} \over \omega_{0}} d \omega_{-}.
\end{split}
\label{localglobal}
\ee

The constraints of section \ref{entanglednull} now lead to the following constraints on these modes. At the outer horizon we have
\be
\label{btzoutside}
\begin{split}
&\left( \anh - e^{-{\pi \omega_0}} \tildanh^{\dagger} \right) |\Psi \rangle = 0; \qquad \left( \anh^{\dagger} - e^{{\pi \omega_0}} \tildanh \right) |\Psi \rangle = 0; \qquad [\anh, \tildanh] = 0;  \\
&\langle \Psi | \anh \anh^{\dagger} | \Psi \rangle = \langle \Psi | \tildanh \tildanh^{\dagger} | \Psi \rangle = {1 \over 1 - e^{-2 \pi \omega_0}}.
\end{split}
\ee
These constraints are automatically met in the Hartle-Hawking state. For the occupation of $\anh$ this follows since, in that state, the global modes are populated as\footnote{Notice that the occupation levels in \eqref{globoclev} know about the temperature of the black hole via the factor $\kappa_+$ which enters in the relation \eqref{defomegaplus} between $\omega_+$  and $\omega$.}
\be
\langle \Psi| a_{\omega_{+}, m} a_{\omega'_{+}, m'}^{\dagger}  | \Psi \rangle =  {1 \over 1 - e^{-2 \pi \omega_{+}}} \delta(\omega_{+} - \omega_{+}') \delta_{m m'}.
\label{globoclev}
\ee
Moreover, the global $\ta$ modes can be constructed using the mirror-operator construction  \cite{Papadodimas:2013jku}, which yields modes whose two-point function is again
\be
\langle \Psi| \ta_{\omega_{+}, m} \ta_{\omega'_{+}, m'}^{\dagger}  | \Psi \rangle =  {1 \over 1 - e^{-2 \pi \omega_{+}}} \delta(\omega_{+} - \omega_{+}') \delta_{m m'},
\ee
and which are moreover entangled with the modes outside the horizon through
\be
\begin{split}
&a_{\omega_{+}, m} |\Psi \rangle = e^{-\pi \omega_{+}} \ta^{\dagger}_{\omega_{+}, m} |\Psi \rangle; \qquad a_{\omega_{+}, m}^{\dagger} |\Psi \rangle = e^{\pi \omega_{+}} \ta_{\omega_{+}, m} |\Psi \rangle; \\
&\ta_{\omega_{+}, m} |\Psi \rangle = e^{-\pi \omega_{+}} \a^{\dagger}_{\omega_{+}, m} |\Psi \rangle; \qquad \ta_{\omega_{+}, m}^{\dagger} |\Psi \rangle = e^{\pi \omega_{+}} a_{\omega_{+}, m} |\Psi \rangle.
\end{split}
\ee
Using \eqref{localglobal}, we can then check that all the relations in \eqref{btzoutside} are satisfied.

Of more interest to us are the constraints that the analysis of section \ref{entanglednull} places on the modes near the inner horizon. Here the constraints of section \ref{entanglednull} can be divided into two parts. The first part is that as we approach the inner horizon from the exterior, the modes should be correctly populated
\be
\label{boccupancy}
\langle \Psi | \bnh \bnh^{\dagger} | \Psi \rangle = {1 \over 1 - e^{-2 \pi \omega_1}}. 
\ee
The second condition is that the modes  behind the inner horizon should be correctly entangled with the modes in front
\be
\label{bcentanglement}
\left( \bnh - e^{-{\pi \omega_1}} \cnh^{\dagger} \right) |\Psi \rangle = 0; \qquad \left( \bnh^{\dagger} - e^{{\pi \omega_1}} \cnh \right) |\Psi \rangle = 0; \qquad [\bnh, \cnh] = 0.
\ee
We check \eqref{boccupancy} in section \ref{frontcheck} and analyze \eqref{bcentanglement} in section \ref{backcheck}.

\subsection{Checking the constraints on approaching the inner horizon \label{frontcheck}}
The constraint \eqref{boccupancy} can be checked by using the propagation of fields between the inner and outer horizon and the propagation of fields from the boundary to the outer horizon. First we note that by virtue of the Killing isometry of the geometry we have
\be
\label{innerouterbog}
\bnh  = \boga \anh  + \bogta \tildanh^{\dagger},
\ee
Here the Bogoliubov coefficients, $\boga$ and $\bogta$ can be obtained from the reflection and transmission coefficients given in \cite{Dias:2019ery}, after accounting for our different normalizations and also accounting for the relationship between local and global modes described. They are given by
\be
\begin{split}
   \boga=  e^{i \delta_1} \left({\kappa_{+} \over \sqrt{2}} \right)^{i \omega_{+}} \frac{\sqrt{\frac{\omega_{-}}{\omega_{+}}} \Gamma (i \omega_{-}) \Gamma \left(1+i \omega_+\right)}{\Gamma
   \left(\frac{1}{2} (-\Delta+i \omega_{-}+i \omega_{+} )+1\right) \Gamma \left(\frac{1}{2} (\Delta + i \omega_{-}+i \omega_{+})\right)} e^{i \delta}
   \\
\bogta =  e^{i \delta_1}  \left({\kappa_{+} \over \sqrt{2}} \right)^{-i \omega_{+}}   \frac{\sqrt{\frac{\omega_{-}}{\omega_{+}}} \Gamma (i \omega_{-}) \Gamma \left(1-i \omega_{+}\right)}{\Gamma \left(\frac{1}{2} (-\Delta +i \omega_{-}-i \omega_{+}) +1\right)\Gamma \left(
\frac{1}{2} (\Delta + i \omega_{-}-i \omega_{+}\right) },
\end{split}
\ee
where the important phase $e^{i \delta}$ is specified in \eqref{Cdeltaformula} and the irrelevant common phase-factor is given by 
\be
e^{i \delta_1} =  \left(\sqrt{2} \rconst \over \kappa_{-} \right)^{i \omega_{-}} {\left(U_0 \kappa_{+} \right)^{i \omega_+ \over \kappa_{+}} \over \left(U_0 \kappa_{-} \right)^{i \omega_- \over \kappa_{-}}} .
\ee

Note that, as we will check below, these Bogoliubov coefficients correctly satisfy
\be
|\boga|^2 - |\bogta|^2 = 1.
\ee
Therefore we find that the condition that must be met for \eqref{boccupancy} to be satisfied is that
\be
|\boga|^2{1 \over 1 - e^{-2\pi \omega_{+}}} + |\bogta|^2 {e^{-2\pi \omega_{+}} \over 1 - e^{-2\pi\omega_{+}}} + \left(\boga \bogta^*  + \bogta \boga^* \right){ e^{-\pi \omega_{+}}  \over 1 - e^{-2\pi \omega_{+}}} = {1 \over 1 - e^{-2\pi \omega_{-}}}.
\label{bogproof}
\ee

At first sight this might seem a little surprising. But, in fact, the identity above is true as can be checked by just repeatedly using the Gamma-function identity $\Gamma(i z) \Gamma(-i z) = {\pi \over z \sinh(\pi z)}$. In particular, we find that
\be
\begin{split}
&\boga \bogta^* = -\frac{1}{2} {1 \over \sinh(\pi  \omega_{-}) \sinh(\pi  \omega_{+})} (\cosh (\pi  \omega_{-})-\cosh (\pi  (\omega_{+}+i \Delta )));  \\
&\boga^* \bogta = -\frac{1}{2} {1 \over \sinh(\pi  \omega_{-}) \sinh(\pi  \omega_{+})} (\cosh (\pi  \omega_{-})-\cosh (\pi  (\omega_{+}-i \Delta ))); \\
&|\boga|^2 = \frac{1}{2} {1 \over \sinh(\pi  \omega_{-}) \sinh (\pi  \omega_{+})} (\cosh (\pi  (\omega_{-}+\omega_{+}))-\cos (\pi  \Delta )); \\
&|\bogta|^2 = \frac{1}{2} {1 \over \sinh(\pi  \omega_{-}) \sinh(\pi  \omega_{+})} (\cosh (\pi  (\omega_{-}-\omega_{+}))-\cos (\pi  \Delta )).
\end{split}
\ee
Putting these results together, with a little algebra, we find that \eqref{bogproof} follows!

We should emphasize that this result arises as a result of a nontrivial conspiracy between the reflection and transmission coefficients that control the propagation between the outer and the inner horizon, and the phase factor $e^{-i \delta}$ that arises from propagation outside the outer horizon. This is reminiscent of the conspiracy between properties of the mode functions in these regions that the authors of \cite{Dias:2019ery} noticed when they were considering the {\em classical} problem in the same background.
 
Our test does not appear to be sensitive to the constraint, ${\Delta r_{-} \over r_{+} - r_{-}} > 1$, that was found to be necessary in \cite{Dias:2019ery} for the stress-tensor to be regular at the inner horizon. Indeed, from our point of view, this constraint is somewhat surprising since \eqref{bogproof} tells us that the state is {\em as smooth as possible} at the inner horizon: all the modes are occupied at just the right temperature. It would be interesting to understand this additional constraint through a mode-sum calculation of the stress-tensor near the inner horizon \cite{Candelas:1980zt}. 

\subsection{Extending the field behind the inner horizon \label{backcheck}}
The second part of our test in section \ref{entanglednull} was that operators in front of the horizon must also be correctly entangled with operators behind it. In this case, the key-point is to focus on the operators $\cnh$.
One might have thought that since the $\cnh$ operators could, in principle, be ``new'' operators, one could just write them down using the standard mirror operator construction \cite{Papadodimas:2013jku}. However, this is not possible due to the monogamy of entanglement. Since the modes just outside the inner horizon are linear combinations of modes near the outer horizon \eqref{innerouterbog}, and since those modes are already entangled with modes outside the outer horizon as in \eqref{localglobal}, \eqref{btzoutside}, the modes near the inner horizon  cannot also be entangled with fresh modes inside the inner horizon. Note that this application of the monogamy of entanglement is very different from the one used in the fuzzball or firewall arguments, since it can be phrased entirely at the level of effective field theory and involves only simple operators. Nevertheless, it turns out to be possible to cleverly  reuse the modes behind the inner and outer horizon to generate modes behind the inner horizon. 

\paragraph{ No new modes \\}
The fact that any new modes behind the inner horizon decouple can be seen from the following argument. Let us write
\be
\label{cnhexpansion}
\cnh = \bogc \tildanh + \bogd \anh^{\dagger} + \boge \dnh + \bogf \enh^{\dagger}.
\ee
where $\dnh$ and $\enh$ denote candidate new modes which commute with $\tildanh$ and $\anh^{\dagger}$.  We will first show that the new modes  $\dnh$ and $\enh^{\dagger}$ can consistently be set to zero above. 

This is done as follows. First, we use the maximal entanglement of $\anh$ and $\tildanh$ to show that the new modes cannot have any correlators with them. We start with
\be
\label{anhannihilating}
\left( \anh - e^{-{\pi \omega_0}} \tildanh^{\dagger} \right) |\Psi \rangle = 0,
\ee
and therefore,
\be
\langle \Psi | \dnh \left( \anh - e^{-\pi \omega_0} \tildanh^{\dagger} \right) |\Psi \rangle = 0.
\ee
But since $\dnh$ computes both with $\anh$ and with $\tildanh^{\dagger}$,  
\be
\langle \Psi |  \left(\anh -  e^{-\pi \omega_0 } \tildanh^{\dagger} \right) \dnh | \Psi \rangle = 0.
\ee
But note that we also have
\be
\langle \Psi| \anh = e^{\pi \omega_0}  \langle \Psi | \tildanh^{\dagger}.
\ee
Substituting this into the previous equation, we find that
\be
2 \sinh \left(\pi \omega_0 \right) \langle \Psi |  \tildanh^{\dagger}  \dnh | \Psi \rangle = 0.
\ee
After employing similar reasoning for the two-point function with $\anh$,
we conclude that
\be
\label{nonewent1}
\langle \Psi| \anh \dnh |\Psi \rangle = \langle \Psi |\tildanh^{\dagger} \dnh | \Psi \rangle = 0.
\ee
Similarly, we have
\be
\label{nonewent2}
\langle \Psi| \anh \enh^{\dagger} |\Psi \rangle = \langle \Psi |\tildanh^{\dagger} \enh^{\dagger} | \Psi \rangle = 0.
\ee

The constraints \eqref{bcentanglement} require
\be
\cnh |\Psi \rangle = e^{-{\pi \omega_1}} \bnh^{\dagger} |\Psi \rangle.
\ee
Expanding both the left and the right hand sides using \eqref{cnhexpansion} and \eqref{innerouterbog}, we see that
\be
\left(\bogc \tildanh + \bogd \anh^{\dagger} + \boge \dnh + \bogf \enh^{\dagger} \right) |\Psi \rangle =  e^{-{\pi \omega_1 }} \left(\boga^* \anh^{\dagger} + \bogta^* \tildanh\right) |\Psi \rangle.
\ee
But since $\dnh |\Psi \rangle$ and $\enh^{\dagger} |\Psi \rangle$  are orthogonal to all  the other vectors that appear above, we see immediately that
\be
\label{nonewads1}
(\boge \dnh + \bogf \enh^{\dagger}) |\Psi \rangle = 0.
\ee
But now using that
\be
\cnh^{\dagger} |\Psi \rangle = e^{\pi \omega_1} \bnh |\Psi \rangle,
\ee
we can also conclude that
\be
\label{nonewads2}
(\boge^* \dnh^{\dagger} + \bogf^* \enh) |\Psi \rangle = 0.
\ee
But from \eqref{nonewads2} and \eqref{nonewads1}, we find that
\be
\label{nocomm}
|\boge|^2 - |\bogf|^2 = 0.
\ee
So not only are the new modes not entangled with the old modes, they cannot even contribute to the ``norm'' of the oscillators behind the horizon!

The results \eqref{nonewent1}, \eqref{nonewent2} and \eqref{nocomm} together constitute an important result. They show us that {\em we cannot define new modes behind the inner horizon that automatically have the right entanglement with modes in front of the inner horizon.} The reader might worry why the mirror operator construction \cite{Papadodimas:2013jku} fails at the inner horizon. The reason is given by the equation \eqref{anhannihilating}. This means that if we look at the set of {\em simple operators} outside the inner horizon, the state is {\em not} separating with respect to these operators. 

Therefore, for the remainder of this analysis, we will set
\be
\boge = \bogf = 0.
\ee

\begin{figure}[!h]
\begin{center}
\includegraphics[width=0.3\textwidth]{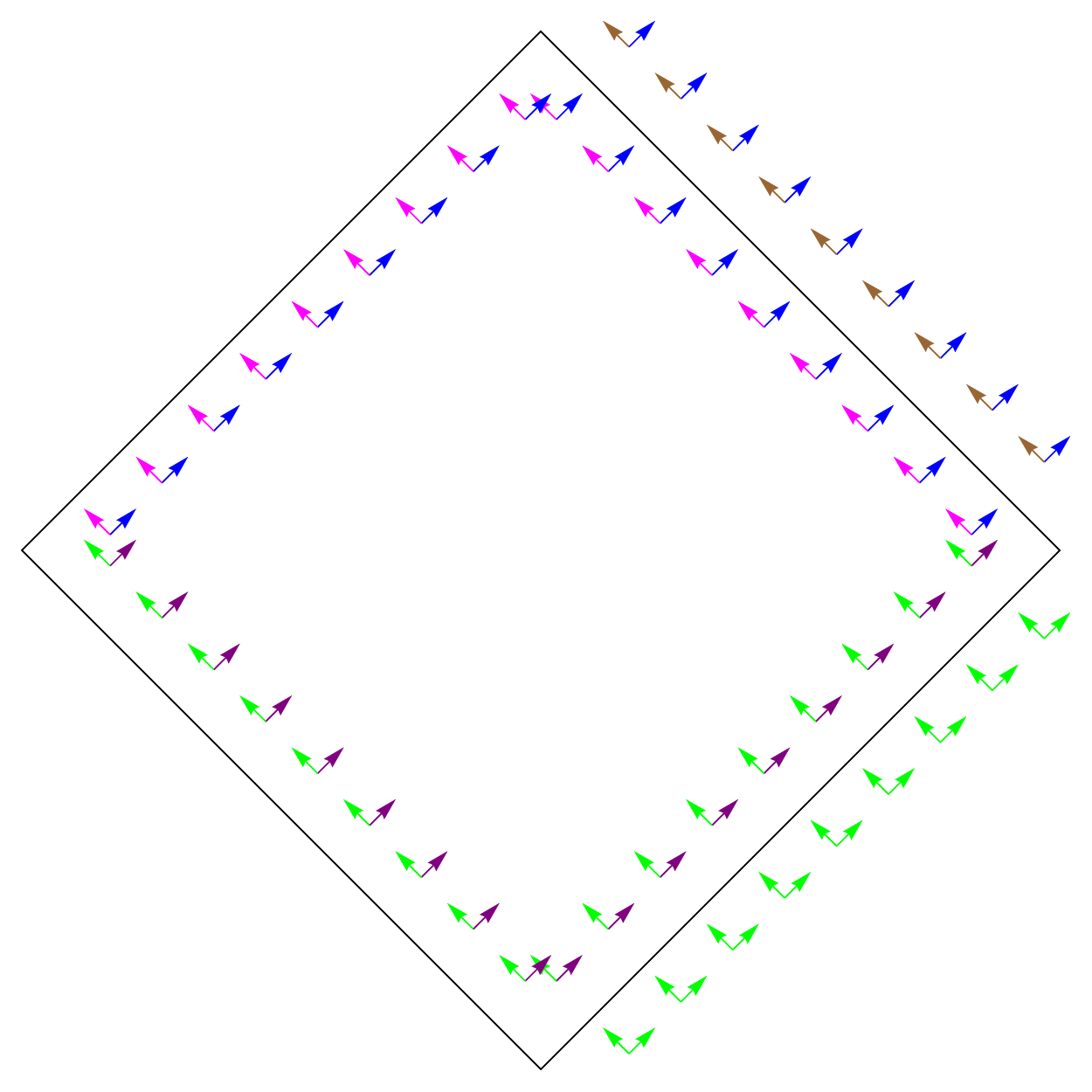}
\caption{\em A figure of the outer and inner horizons and the modes near them. The purple modes near the right inner horizon are linear combinations of the green and violet modes near the outer horizons. But the green and violet modes are already entangled together.   So, it is nontrivial to find the brown modes behind the inner horizon with the correct entanglement.\label{leftrightmovers}}
\end{center}
\end{figure}

A pictorial representation of our result that one cannot arbitrarily define new modes behind the inner horizon is shown in Figure \ref{leftrightmovers}

\paragraph{Reusing old modes \\}
At first the reader might conclude that the analysis above implies that it is impossible to extend the field behind the inner horizon. However, it turns out that is  possible to cleverly {\em reuse} the old modes by choosing appropriate values for $\bogc$ and $\bogd$ in \eqref{cnhexpansion} so as to satisfy the entanglement constraints. 

The concern is that the constraints might overdetermine $\bogc$ and $\bogd$. However, the idea is to use the entanglement constraints between $\cnh$ and $\bnh$ to solve for $\bogc$ and $\bogd$ and then simply check whether $[\cnh,\bnh] = 0$ and $[\cnh, \cnh^{\dagger}] = 1$. This is done as follows.
We first note using \eqref{cnhexpansion} (and setting $\boge = \bogf = 0$)
\be
\cnh |\Psi \rangle = (\bogc \tildanh + \bogd \anh^{\dagger}) |\Psi \rangle = (\bogc + \bogd e^{\pi \omega_0 } ) \tildanh |\Psi \rangle,
\ee
where we have used the entanglement between $\anh$ and $\tildanh$. 
On the other hand, using the entanglement that is required between $\cnh$ and $\bnh$ and using \eqref{innerouterbog} we find that
\be
\begin{split}
\cnh |\Psi \rangle &= e^{-{\pi \omega_1 }} \bnh^{\dagger} | \Psi \rangle = e^{-{\pi \omega_1 }} (\boga^* \anh^{\dagger} + \bogta^*  \tildanh) |\Psi \rangle  \\
&=e^{-{\pi \omega_1 }}  (\boga^* e^{\pi \omega_0 } + \bogta^*)\tildanh |\Psi \rangle,
\end{split}
\ee
where, in the last line, we again used the entanglement between the $\anh$ and the $\tildanh$ modes. Equating the final expressions on the two sides of the equations above, and using the fact that $\tildanh |\Psi \rangle \neq 0$ leads to the condition that
\be
\label{cdab1}
\bogc + \bogd e^{\pi \omega_0 } = e^{-{\pi \omega_1 }}   (\boga^* e^{\pi \omega_0 } + \bogta^*).
\ee
We can perform a similar analysis using the action of $\cnh^{\dagger}$. Here we find that
\be
\cnh^{\dagger} |\Psi \rangle = (\bogc^* \tildanh^{\dagger} + \bogd^* \anh) |\Psi \rangle = (\bogc^* + \bogd^* e^{-\pi \omega_0 } ) \tildanh^{\dagger} |\Psi \rangle. 
\ee
On the other hand, the smoothness of the inner horizon requires
\be
\begin{split}
\cnh^{\dagger} |\Psi \rangle &= e^{{\pi \omega_1 }} \bnh | \Psi \rangle  = e^{\pi \omega_1 }  (\bogta \tildanh^{\dagger} + \boga \anh) |\Psi \rangle  \\
&= e^{\pi \omega_{1} }  \left( \boga e^{-\pi \omega_0 } + \bogta \right) \tildanh^{\dagger} | \Psi \rangle.
\end{split}
\ee
Taking the coefficients of the right hand sides of the two results above, and then taking the complex conjugate leads to the relation
\be
\label{cdab2}
\bogc + \bogd e^{-\pi \omega_0 } = e^{\pi \omega_1}  \left(\boga^* e^{-\pi \omega_0 } + \bogta^* \right).
\ee

The relations \eqref{cdab1} and \eqref{cdab2} give linear equations for $\bogc$ and $\bogd$ that can be solved in terms of $\boga,\bogta$. In particular the solution is
\be
\begin{split}
&\bogc = {1 \over \sinh(\pi \omega_0)} \left[ \boga^* \,\sinh(\pi \omega_1)+ \bogta^* \,\sinh[\pi(\omega_0+\omega_1)]\right]; \\
&\bogd = {1 \over \sinh(\pi \omega_0)}\left[ \boga^* \, \sinh[\pi(\omega_0-\omega_1)]- \bogta^* \,\sinh(\pi \omega_1)\right].
\end{split}
\ee
These solutions are subject to further consistency checks. The first consistency check is that we need $[\cnh, \bnh] = 0$. This leads to the requirement
\be
\label{zerocommcb}
\boga \bogd - \bogta \bogc = 0.
\ee
Using \eqref{cdab1} and \eqref{cdab2} we find, after some algebra, that
\be
\boga \bogd - \bogta \bogc = {1\over 2}{ e^{-\pi \omega_1} \over \coth(\pi \omega_0)-1}\left(|\boga e^{\pi \omega_0} + \bogta|^2  - |\boga + \bogta e^{\pi \omega_0}|^2 e^{2 \pi \omega_1} \right).
\label{abcd}
\ee
On the other hand, from the relations
\be
\langle \Psi | \bnh \bnh^{\dagger} | \Psi \rangle = 1 + \langle \Psi | \bnh^{\dagger} \bnh | \Psi \rangle =  { 1 \over 1 - e^{-2 \pi \omega_1}},
\ee
we find that
\be
|\boga + \bogta e^{\pi \omega_0 }|^2 = e^{2 \pi (\omega_0 - \omega_1)} {1 - e^{-2 \pi \omega_0} \over 1 - e^{-2 \pi \omega_1}},
\ee
and also that
\be
|\boga + \bogta e^{-\pi \omega_0}|^2 = {1 - e^{-2 \pi \omega_0} \over 1 - e^{-2 \pi \omega_1}}.
\ee
Substituting these relations in \eqref{abcd} we find that indeed \eqref{zerocommcb} is satisfied.

The second consistency check is that the solution must satisfy $[\cnh, \cnh^{\dagger}] = 1$. This is just the requirement that 
\be
|\bogc|^2 - |\bogd|^2 = 1.
\ee
But this is also satisfied since 
\be
|\bogc|^2 - |\bogd|^2 = { e^{-2 \pi \omega_1}  \over (e^{2 \pi \omega_0} - 1)}\left[-|\boga e^{\pi \omega_0}+\bogta  |^2  + |\boga + \bogta e^{\pi \omega_0}|^2 e^{4 \pi \omega_0}\right] = 1.
\ee

One might have hoped that by placing reflecting boundary conditions at the singularity, one would correctly reproduce the coefficients $\bogc$ and $\bogd$ i.e. we quantize the field behind the inner horizon assuming that the geometry is given the naive analytic continuation and then also impose $\phi = 0$ at the timelike singularity at $r = 0$. However, a simple calculation shows that this does {\em not} give the right values of $\bogc$ and $\bogd$. It is possible that more sophisticated reflecting boundary conditions could be used to reproduce the values of $\bogc$ and $\bogd$ but we have not yet discovered them.

\section{Conclusions and discussion \label{secdiscussion}}
\paragraph{ Summary of results. }
In this paper, we have developed a new test that provides a necessary condition for a quantum state to be smooth in the vicinity of a null surface. The condition is that the near-horizon modes defined by \eqref{defnearhormodes}, which have canonical commutators, must have the self-correlators given by \eqref{universalself} and must be entangled with each other through the correlators \eqref{universalentang}. This is a universal result, which relies just on the short-distance properties of the correlator of the scalar field.

When one considers static black holes, where the field can also be expanded in global modes that are obtained as Fourier transforms of the field, then 
this test can be translated into the statement that the two-point correlator of such global modes has a delta-function piece in the frequency-difference, and the coefficient of this delta function is the same in any smooth state.

We applied this test to charged black holes in anti-de Sitter space for boundary dimension, $3,4,5,6$, where we were able to easily verify, for a range of parameters, that the naturally defined Hartle-Hawking state was not smooth on the inner horizon. So, even in the absence of external perturbations, quantum fluctuations destabilize the inner horizon for such black holes. These results can be understood to be the AdS analogue of the results of \cite{Sela:2018xko} for charged flat-space black holes in four dimensions but they also demonstrate the utility of our test.

We then considered the BTZ black hole, which was recently argued to violate strong cosmic censorship in \cite{Dias:2019ery}. Here, we found, that part of our test --- which pertained to the occupation number of modes as we approach the inner horizon ---  was automatically satisfied in the Hartle-Hawking state as a result of some non-trivial identities involving the various reflection and transmission coefficients in the black-hole geometry. 

However, our test has a second element: it requires the existence of modes behind the inner horizon that are correctly entangled with modes in front. This presents an additional complication: although, one might naively assume that one could  expand the field in terms of new degrees of freedom behind the inner horizon, this turns out to be prohibited by the monogamy of entanglement. Instead, one is forced to reuse the degrees of freedom between the inner and outer horizon to construct near-horizon modes behind the inner horizon. We identified the correct combination that could be used to expand the field just behind the inner horizon.

We clarify that this does {\em not} completely fix the dynamics behind the inner horizon. If one considers points that are finitely separated from the inner horizon, then the correlators of field insertions at such points require information that is {\em not} contained in the near-horizon modes. The arbitrariness in these correlators corresponds to the freedom that one has in extending the metric and the fields behind the Cauchy horizon. 

Our results seem to support the claim that strong cosmic censorship is violated for the BTZ black hole. If so, the arbitrariness described above would present interesting challenges for the AdS/CFT interpretation of such black holes. The AdS/CFT conjecture suggests that a boundary CFT describes all phenomenon in the bulk geometry but a traversable inner horizon would imply that there are events (behind the inner horizon) that cannot be described by the boundary CFT in an obvious way. While this would be an interesting scenario, we believe that it is premature to conclude that strong cosmic censorship is violated for the BTZ black hole. 

\paragraph{ Unresolved questions about strong cosmic censorship.}
Our reasons for believing that strong cosmic censorship continues to be a subtle issue, at the quantum level,  are as follows. First, we emphasize that from the point of view of the exterior, strong cosmic censorship is a statement about the infinite future, since any events at the inner horizon are to the future of all events outside the outer horizon. However, at very late times, we expect a host of ill-understood and nonperturbative phenomena in the black-hole exterior.
   
The first such phenomena is that, even for a fixed angular momentum,  perturbations outside the black hole do {\em not} die off exponentially for arbitrarily long times. Instead at a time of $\Or[S]$ where $S$ is the black hole entropy, the perturbation reaches a size $e^{-{S \over 2}}$ and develops a fat tail, where it does not decay any further \cite{Maldacena:2001kr}. What does this fat tail imply for strong cosmic censorship? Even the classical arguments in favour of the violation of strong cosmic censorship rely on a careful balance between the blue-shift near the inner horizon and the decay outside. If even an exponentially small amount of energy falls into the black hole at very late times, this could destabilize the inner horizon.
 
At even later times $\Or[e^{e^{S}}]$ there are even more exotic phenomena. Since the BTZ black hole is believed to be dual to a theory with a discrete spectrum of states, at very long times, we expect the entire system to under Poincare recurrence. Thus, if the black hole was formed from collapse by matter thrown in from the boundary, then Poincare recurrence implies that that the black hole should eventually spontaneously turn into a white hole that emits matter, which then proceeds to bounce off the boundary, and re-collapse into the black hole. What does this imply for the inner horizon? 

These nonperturbative effects are ill understood and not addressed by our effective field theory analysis. Nevertheless, they are crucial to address the issue of strong cosmic censorship precisely because this analysis poses peculiar questions about events in the infinite future.

\paragraph{ Discussion.}
Our work points to several directions for future work. One technical issue, which we believe would be interesting to understand better, is as follows. In the Hartle-Hawking state, the occupancy of the modes near the inner horizon of the BTZ black hole, which we presented in section \ref{secbtz}, provides enough information to compute the stress-tensor at the inner horizon using mode sums. It would be nice to understand the constraint found in \cite{Dias:2019ery}--- which suggests that the stress-tensor is finite only $\beta = {\Delta \over {r_+ \over r_{-} - 1}} > 1$ --- from the perspective of direct mode sums.

It is also natural to use our test to understand both flat-space and de Sitter black holes. The de Sitter case is particularly interesting because, classically, not only do charged de-Sitter black holes violate even the weak formulations of strong cosmic censorship \cite{Dias:2018ynt}, even non rotating and neutral black holes in de Sitter space have two horizons: the black-hole and the cosmological horizon. So an interesting exercise is to consider various possible quantum states and examine their smoothness on all these horizon by the methods discussed in this paper.  We hope to return to this analysis in forthcoming work.

{\bf Acknowledgments:} We would like to thank Jose Barbon, Jan de Boer, Chandramouli Chowdhury, Alessandra Gnecchi, Monica Guica,  Chandan Jana, R. Loganayagam,  Shiraz Minwalla, Ruchira Mishra, K. Narayan,  Harvey Reall, Erik Verlinde and Amitabh Virmani for helpful discussions. PS is grateful to TIFR (Mumbai) for hospitality while this work was being completed. The work of SR is supported, in part,  by a Swarnajayanti fellowship of the Department of Science and Technology (India).

\appendix
\section*{Appendix}
\label{applargel}
\section{Scattering in the Reissner-Nordstr\"{o}m geometry at large angular momenta  \label{appendixwkb}}

In Section \ref{secnumerics}, we noticed that the fractional difference between the expected Boltzmann factor and the numerical Boltzmann factor at the inner horizon for large $\ell$ modes in  Reissner-Nordstr\"{o}m black holes tends to zero. Here, we will explain this feature by using WKB approximation to solve wave the equation at large $\ell$. The exact form of the metric is given in Section \ref{secrnads}.
\be
ds^2 = -f(r) dt^2 + f(r)^{-1} dr^2 + r^2 d\Omega^2_{d-1}.
\ee
We reproduce the near horizon form of $f(r)$.
\be
f(r) = 2 \kappa_{\pm} (r-r_{\pm}) ; \qquad r\rightarrow r_{\pm}.
\ee
We consider a massless scalar Klein-Gordon field $\phi$.
The wave equation can be solved with an ansatz of the form
\be
\phi_{\omega,\ell}(r,t,\Omega) = {1 \over r^{(d-1)/2}} \psi_{\ell,\omega}(r_*) e^{-\mathnormal{i}\omega t} Y_\ell^m(\Omega),
\ee
which leads to
\be\label{nohweqn}
\begin{split} 
&\psi_{\ell,\omega}''(r_*) - V(r_*) \psi_{\ell,\omega}(r_*) = 0, \\
& V(r_*) = -\omega^2 + {f(r) \over r^2} \left(\ell(\ell+d-2) +\frac{(d-3)(d-1)}{4} f(r) + {(d-1) \over 2} rf'(r)\right),
\end{split} 
\ee
where $r_*$ is the \emph{tortoise} coordinate, defined by $dr_* = {dr\over f(r)}$
\subsubsection*{Large-$\ell$ behaviour}
At large $\ell$, the potential near outer horizon takes the following form.
\be
V(r_*) \approx - \omega^2 + {2\kappa_+ (r-r_+) \over r_+^2}\ell^2.
\ee
Solving the $\psi$-equation, \eq{nohweqn}, with the above potential yields,
\be
\begin{split}
\psi_{\ell,\omega}(r_*) &\approx A_B K_{\mathnormal{i}\omega/\kappa_+}\left({\ell \over r_+\kappa_+} \sqrt{f(r)}\right) + B_B I_{\mathnormal{i}\omega/\kappa_+}\left({\ell \over r_+ \kappa_+} \sqrt{f(r)}\right).
\end{split}
\ee
In the regime, $\ell (r-r_+) \gg 1$ and $r-r_+ \ll r_+$, this solution can be approximated as
\be
A_B \sqrt{{\pi \over 2}} \left({r_+\kappa_+ \over \ell \sqrt{f(r)}}\right)^{1\over2} e^{-{\ell \sqrt{f(r)}\over r_+\kappa_+} } + B_B \sqrt{{1 \over 2\pi}} \left({r_+\kappa_+ \over \ell \sqrt{f(r)}}\right)^{1\over2}  \left[ i e^{-\pi\omega} e^{-{\ell \sqrt{f(r)}\over r_+\kappa_+} } + e^{{\ell \sqrt{f(r)}\over r_+\kappa_+} }\right].
\ee
The near horizon Bessel solution can be matched with a WKB solution in the large-$\ell$ limit. For large-$\ell$, the WKB approximation is valid at all radial points where $f(r)$ is finite (and nonzero). 
\be
\psi_{\text{wkb}}(r_*) \approx A_{\text{wkb}} e^{-\int^{r_*} \sqrt{V} dr_*} + B_{\text{wkb}} e^{\int^{r_*} \sqrt{V} dr_*}.
\ee
Near the boundary,
\be\label{wkblargelboundary}
\psi_{\text{wkb}}(r_*) \approx A_{\text{wkb}}  e^{-{\ell r_*}} + B_{\text{wkb}}  e^{{\ell r_*} } , \qquad 1\ll r\ll \ell.
\ee
Near the outer horizon, 
\be\label{wkblargeloutsideouterhorizon}
\psi(r_*) \approx A_{\text{wkb}} e^X e^{-{\ell \over r_+\kappa_+} \sqrt{f(r)}} + B_{\text{wkb}} e^{-X} e^{{\ell \over r_+\kappa_+} \sqrt{f(r)}} , \qquad {1\over\ell}\ll r-r_+\ll r_+,
\ee
where, \be 
 X = {\int_{\zeta^+_*}^{\zeta^{\infty}_*}\sqrt{V(r)} dr_*}   \ee
with ${1\over\ell}\ll \zeta_+ -r_+\ll r_+$ and $1 \ll \zeta_{\infty}\ll \ell$ close to boundary. To a very good approximation, we can write
\be
X \approx l \int_{-\infty}^{0} {\sqrt{f(r)} \over r} dr_* \equiv l \Lambda , \qquad \Lambda>0 .
\ee
Near the boundary, the normalizable mode behaves as
\be
\psi(r_*) \approx N \sqrt{r_*} J_{d\over 2}\left( -\mathnormal{i}r_*\sqrt{\ell^2-\omega^2}\right) , \qquad r\gg \ell,
\ee
where $N$ is a normalization constant. For large-$\ell$, the normalizable mode behaves as
\be
\psi(r_*) \approx N {\mathnormal{i}\over\sqrt{\ell}}{1\over \sqrt{2\pi}}\left(e^{- \ell r_* } e^{{\mathnormal{i} \pi\over 8}d} + \mathnormal{i} e^{l r_*}e^{{3\mathnormal{i} \pi\over 8}d}\right) , \qquad 1 \ll r \ll \ell.
\ee 
Matching this with the WKB solution \eq{wkblargelboundary}, we get
\be
{A_{\text{wkb}} \over B_{\text{wkb}}} = -\mathnormal{i} e^{-{\mathnormal{i} \pi\over 4}d}.
\ee
Near horizon behaviour would now become
\be
\psi(r_*) \approx N {\mathnormal{i}\over\sqrt{\ell}}{1\over \sqrt{2\pi}} \left( -\mathnormal{i} e^{-{\mathnormal{i} \pi\over 4}d} e^{l\Lambda}  e^{-{\ell \over r_+\kappa_+} \sqrt{f(r)}} + e^{-l\Lambda} e^{{\ell \over r_+\kappa_+} \sqrt{f(r)}}\right).
\ee
We see that the coefficient of $e^{{\ell \over r_+\kappa_+} \sqrt{f(r)}}$ is exponentially suppressed. Hence, in the large $\ell$ limit,
\be 
B_B = 0
\ee
Close to horizon,
\be 
 \psi_{\ell,\omega}(r_*) = A_B K_{\mathnormal{i}\omega/\kappa_+}\left({\ell \over r_+\kappa_+} \sqrt{f(r)}\right)\ee
Now we consider the near horizon ($\ell (r-r_+) \ll 1$) limit.
\be
\begin{split}
\psi_{\ell,\omega}(r_*) = & \frac{A_B}{2} \left[ \left({\ell\over 2 r_+ \kappa_+}\right)^{\mathnormal{i}\omega \over \kappa_+} \Gamma(-\mathnormal{i \omega \over \kappa_+}) e^{\mathnormal{i} \omega r_*}  + (\omega\rightarrow-\omega)\right] \\
=& \frac{A_B}{2} \left({\pi \kappa_+\over \omega \sinh({\pi \omega \over \kappa_+})}\right)^{1/2}\left[ e^{\mathnormal{i}\delta_{\omega,\ell} \over 2 } e^{\mathnormal{i} \omega r_*}  + e^{-\mathnormal{i}\delta_{\omega,\ell}\over 2}e^{-\mathnormal{i}\omega r_*}\right],
\end{split}
\ee
where
\be\label{phaselargel}
{\delta_{\omega,\ell} \over 2} = \arg\left(\left({\ell\over 2 r_+\kappa_+}\right)^{\mathnormal{i}\omega \over \kappa_+} \Gamma(-\mathnormal{i \omega \over \kappa_+})\right) ; \hspace{20 pt} \text{large-}\ell.
\ee
The mode expansion of the scalar field just outside the horizon as $r \rightarrow r_+$ is 
\be
\phi(r,t,\Omega) = \int {d\omega \over 2\pi} {1\over\sqrt{2\omega}} {1\over r_+^{d-1\over2}}\sum_{\ell,m} e^{-\mathnormal{i}\omega t} Y_{\ell,m}(\Omega) a_{\omega,\ell}\left[ e^{-i \delta_{\omega,\ell}}e^{-\mathnormal{i}\omega r_*} + e^{i \delta_{\omega,\ell} } e^{\mathnormal{i}\omega r_*} \right] + \hc
\ee
\subsubsection*{Scattering between horizons for large $\ell$}
The mode expansion just inside the outer horizon, as $r \rightarrow r_+ + 0^-$ is as follows.
\be
\phi(r,t,\Omega) = \int {d\omega \over 2\pi} {1\over\sqrt{2\omega}}  {1\over r_+^{d-1\over2}}\sum_{\ell,m} e^{-\mathnormal{i}\omega r_*} \left[e^{-    \mathnormal{i}\delta_{\omega,\ell}} a_{\omega,\ell} e^{-\mathnormal{i}\omega t} Y_{\ell,m}(\Omega) +  \tilde{a}_{\omega,\ell} e^{\mathnormal{i}\omega t} \bar{Y}_{\ell,m}(\Omega)\right] + \hc.
\ee
While, just outside the inner horizon, as $r \rightarrow r_- + 0^+$,
\be
\phi(r,t,\Omega) = \int {d\omega \over 2\pi} {1\over\sqrt{2\omega}}  {1\over r_-^{d-1\over2}}\sum_{\ell,m} e^{-\mathnormal{i}\omega r_*} \left[b_{\omega,\ell} e^{-\mathnormal{i}\omega t} Y_{\ell,m}(\Omega) +  \tilde{b}_{\omega,\ell} e^{\mathnormal{i}\omega t} \bar{Y}_{\ell,m}(\Omega)\right] + \hc
\ee
To compute the occupation numbers of the modes at the inner horizon, we need to relate the operators $a$ and $b$. Consider a solution, $Z(r)$, to the wave equation such that
\be
Z(r) \approx \Gamma(1+\mathnormal{i}\frac{\omega}{\kappa_-} ) \left({\ell \zeta \over 2 r_- \kappa_-}\right)^{-\mathnormal{i}{\omega\over\kappa_-}}  J_{\mathnormal{i}{\omega\over\kappa_-}} \left(\frac{\ell}{r_- \kappa_-} \sqrt{|f(r)|} \right) ; \hspace{20 pt} r-r_- \ll r_-,
 \ee 
 where $\zeta$ is the factor introduced in the definition of the tortoise coordinate \eqref{rtorrinner}.
Close to the inner horizon,
\be
Z(r) = e^{-\mathnormal{i}\omega r_*} ; \hspace{20 pt} r\rightarrow r_-+0^+
\ee 
When we move slightly away from the inner horizon such that $\ell (r-r_-) \gg 1$ but $r-r_- \ll r_-$,
\be
\begin{split}
Z(r) \approx &\Gamma(1+\mathnormal{i}\frac{\omega}{\kappa_-} ) \left({\ell \zeta\over 2 r_- \kappa_-}\right)^{- \mathnormal{i}{\omega\over\kappa_-}}   \left( \frac{r_- \kappa_-}{2\pi\ell \sqrt{|f(r)|}}    \right)^{1\over2} \\ &\times \left[ e^{-{\mathnormal{i}\pi\over 4}} e^{\pi \omega\over 2\kappa_-} e^{\mathnormal{i}  \left(\frac{\ell}{r_- \kappa_-} \sqrt{|f(r)|} \right)}  +  e^{\mathnormal{i}\pi\over 4} e^{-{\pi \omega\over 2\kappa_-}} e^{-\mathnormal{i} \left(\frac{\ell}{r_- \kappa_-} \sqrt{|f(r)|} \right)} \right].
\end{split}
\ee 
Away from the horizon, we can also use the large-$\ell$ WKB approximation.
\be
\begin{split}
Z(r) &\approx {1\over V^{1/4}}\left[A_+ e^{\mathnormal{i} \int_{ -\infty}^{r_*} \sqrt{V} dr_* } + B_+ e^{-\mathnormal{i} \int_{ -\infty}^{r_*} \sqrt{V} dr_* }\right]\\
Z(r) &\approx {1\over V^{1/4}}\left[A_- e^{\mathnormal{i} \int_{ \infty}^{r_*} \sqrt{V} dr_* } + B_- e^{-\mathnormal{i} \int_{ \infty}^{r_*} \sqrt{V} dr_* }\right],
\end{split}
\ee
with, 
\be
 A_- = A_+  e^{\mathnormal{i} \int_{- \infty}^{\infty} \sqrt{V} dr_* } = A_+e^{\mathnormal{i}\theta} \qquad B_- = B_+  e^{-\mathnormal{i} \int_{- \infty}^{\infty} \sqrt{V} dr_* } = B_+e^{-\mathnormal{i}\theta}.
\ee
Matching the WKB solution with the Bessel expansion near the inner horizon, we get
\be
\begin{split}
A_- &= \Gamma(1+\mathnormal{i}\frac{\omega}{\kappa_-} ) \left({\ell \zeta \over 2 r_- \kappa_-}\right)^{-\mathnormal{i}{\omega\over\kappa_-}}  \sqrt{\kappa_-\over2\pi} e^{\mathnormal{i}\pi\over2}e^{-{\pi\omega\over2\kappa_-}} ;\\
B_- &= \Gamma(1+\mathnormal{i}\frac{\omega}{\kappa_-} ) \left({\ell \zeta \over 2 r_-\kappa_-}\right)^{-\mathnormal{i}{\omega\over\kappa_-}}  \sqrt{\kappa_-\over2\pi} e^{\pi\omega\over2\kappa_-}.
\end{split}
\ee
Close to the outer horizon, $\ell (r_+-r) \gg 1$ but $r_+-r \ll r_+$, this solution can also be written as 
\be
\begin{split}
Z(r) \approx \sqrt{2\pi\over\kappa_+} {e^{\pi\omega\over2\kappa_+}\over e^{2\pi\omega\over\kappa_+} -1} \left[ (A_+ e^{\mathnormal{i}{\pi\over4}}  - B_+ e^{-\mathnormal{i}{\pi\over4}} e^{\pi\omega\over\kappa_+} )  J_{-\mathnormal{i}{\omega\over\kappa_+}} \left(\frac{\ell}{r_+ \kappa_+} \sqrt{|f(r)|} \right) \right. \\
\left. + (B_+ e^{-\mathnormal{i}{\pi\over4}}-A_+ e^{\mathnormal{i}{\pi\over4}}e^{\pi\omega\over\kappa_+}  )  J_{\mathnormal{i}{\omega\over\kappa_+}} \left(\frac{\ell}{r_+ \kappa_+} \sqrt{|f(r)|} \right) \right].
\end{split}
\ee 

At the outer horizon, we can expand the Bessel function and substitute for $A_+$ and $B_+$ to get,
\be
Z(r) = A_{\omega,\ell} e^{-i \omega r_*} +B_{\omega,\ell} e^{i \omega r_*} ; \hspace{20 pt} r\rightarrow r_+-0^+,
\ee
where, for large-$\ell$,
\be\label{largelbogoliubov}
\begin{split}
A_{\omega,\ell} &= N_{\omega,\ell} \left[e^{-{\pi\omega\over2\kappa_-}} e^{\mathnormal{i}({\pi\over2}-\theta)}  - e^{\pi\omega\over2\kappa_-} e^{-\mathnormal{i}({\pi\over2}-\theta)} e^{\pi\omega\over\kappa_+} \right]e^{-i{\delta_{\omega,l}\over 2}};\\
B_{\omega,\ell} &= N_{\omega,\ell} \left[e^{-{\pi\omega\over2\kappa_-}} e^{\mathnormal{i}({\pi\over2}-\theta)}e^{\pi\omega\over\kappa_+}-e^{\pi\omega\over2\kappa_-} e^{-\mathnormal{i}({\pi\over2}-\theta)} \right] e^{\mathnormal{i} {\delta_{\omega,l}\over 2}}; \\
N_{\omega,\ell} &= e^{\mathnormal{i}\delta'_{\omega,\ell}} \sqrt{\sinh(\pi{\omega\over\kappa_+}) \over\sinh(\pi{\omega\over\kappa_-})} {e^{\pi\omega\over2\kappa_+}\over e^{2\pi\omega\over\kappa_+} -1}; \\
e^{\mathnormal{i}\delta'_{\omega,\ell}} &= e^{i 3\pi \over 4}\left( {\ell \zeta \over 2 r_- \kappa-}\right)^{-i {\omega \over \kappa_-}} \sqrt{\Gamma\left(1+i {\omega \over \kappa-}\right)\over\Gamma\left(1-i {\omega \over \kappa-}\right) }.
\end{split}
\ee
We can check that the Bogoliubov coefficients satisfy, 
\be
|A_{\omega,\ell}|^2-|B_{\omega,\ell}|^2 = 1.
\ee
Using the mode expansion near the horizons,
\be
b_{\omega,\ell} 
= e^{-i \delta_{\omega,\ell}}A^*_{\omega,\ell} a_{\omega,\ell} - B^*_{\omega,\ell}  \tilde{a}^{\dagger}_{\omega,\ell}.
\ee
Using the constraints on operators $a$ and $\tilde{a}$, \eq{globalconstraints}, and large-$\ell$ Bogoliubov coefficients, \eq{largelbogoliubov}, we get the Boltzmann factor at inner horizon.
\be
\langle b_{\omega,\ell}b^{\dagger}_{\omega',\ell'}\rangle = {1\over 1- e^{-{2\pi \omega \over \kappa_-}}} \delta(\omega-\omega')\delta_{\ell,\ell'} \hspace{20pt} \ell \gg 1.
\ee

\bibliographystyle{JHEP}
\bibliography{references}

\providecommand{\href}[2]{#2}\begingroup\raggedright\begin{thebibliography}{10}

\bibitem{penrose1968structure}
R.~Penrose, \emph{Structure of space-time},  in \emph{Battelle Rencontres: 1967
  lectures in mathematics and physics} (C.~de~Witt and J.~Wheeler, eds.).
\newblock New York: Benjamin, 1968.

\bibitem{Simpson:1973ua}
M.~Simpson and R.~Penrose, \emph{{Internal instability in a Reissner-Nordstrom
  black hole}}, \href{http://dx.doi.org/10.1007/BF00792069}{\emph{Int. J.
  Theor. Phys.} {\bf 7} (1973) 183--197}.

\bibitem{mcnamara1978instability}
J.~M. McNamara, \emph{Instability of black hole inner horizons},
  {\emph{Proceedings of the Royal Society of London. A. Mathematical and
  Physical Sciences} {\bf 358} (1978) 499--517}.

\bibitem{chandrasekhar1982crossing}
S.~Chandrasekhar and J.~B. Hartle, \emph{On crossing the cauchy horizon of a
  reissner--nordstr{\"o}m black-hole}, {\emph{Proceedings of the Royal Society
  of London. A. Mathematical and Physical Sciences} {\bf 384} (1982) 301--315}.

\bibitem{Poisson:1989zz}
E.~Poisson and W.~Israel, \emph{{Inner-horizon instability and mass inflation
  in black holes}},
  \href{http://dx.doi.org/10.1103/PhysRevLett.63.1663}{\emph{Phys. Rev. Lett.}
  {\bf 63} (1989) 1663--1666}.

\bibitem{Poisson:1990eh}
E.~Poisson and W.~Israel, \emph{{Internal structure of black holes}},
  \href{http://dx.doi.org/10.1103/PhysRevD.41.1796}{\emph{Phys. Rev.} {\bf D41}
  (1990) 1796--1809}.

\bibitem{Ori:1991zz}
A.~Ori, \emph{{Inner structure of a charged black hole: An exact mass-inflation
  solution}}, \href{http://dx.doi.org/10.1103/PhysRevLett.67.789}{\emph{Phys.
  Rev. Lett.} {\bf 67} (1991) 789--792}.

\bibitem{Ori:1992zz}
A.~Ori, \emph{{Structure of the singularity inside a realistic rotating black
  hole}}, \href{http://dx.doi.org/10.1103/PhysRevLett.68.2117}{\emph{Phys. Rev.
  Lett.} {\bf 68} (1992) 2117--2120}.

\bibitem{Mellor:1989ac}
F.~Mellor and I.~Moss, \emph{{Stability of Black Holes in De Sitter Space}},
  \href{http://dx.doi.org/10.1103/PhysRevD.41.403}{\emph{Phys. Rev.} {\bf D41}
  (1990) 403}.

\bibitem{Chambers:1994ap}
C.~M. Chambers and I.~G. Moss, \emph{{Stability of the Cauchy horizon in
  Kerr-de Sitter space-times}},
  \href{http://dx.doi.org/10.1088/0264-9381/11/4/019}{\emph{Class. Quant.
  Grav.} {\bf 11} (1994) 1035--1054},
  [\href{http://arxiv.org/abs/gr-qc/9404015}{{\tt gr-qc/9404015}}].

\bibitem{Brady:1996za}
P.~R. Brady, C.~M. Chambers, W.~Krivan and P.~Laguna, \emph{{Telling tails in
  the presence of a cosmological constant}},
  \href{http://dx.doi.org/10.1103/PhysRevD.55.7538}{\emph{Phys. Rev.} {\bf D55}
  (1997) 7538--7545}, [\href{http://arxiv.org/abs/gr-qc/9611056}{{\tt
  gr-qc/9611056}}].

\bibitem{Dafermos:2003wr}
M.~Dafermos, \emph{{The Interior of charged black holes and the problem of
  uniqueness in general relativity}}, {\emph{Commun. Pure Appl. Math.} {\bf 58}
  (2005) 0445--0504}, [\href{http://arxiv.org/abs/gr-qc/0307013}{{\tt
  gr-qc/0307013}}].

\bibitem{Dafermos:2002ka}
M.~Dafermos, \emph{{Stability and instability of the Reissner-Nordstrom Cauchy
  horizon and the problem of uniqueness in general relativity}},
  {\emph{Contemp. Math.} {\bf 350} (2004) 99--113},
  [\href{http://arxiv.org/abs/gr-qc/0209052}{{\tt gr-qc/0209052}}].

\bibitem{Murata:2013daa}
K.~Murata, H.~S. Reall and N.~Tanahashi, \emph{{What happens at the horizon(s)
  of an extreme black hole?}},
  \href{http://dx.doi.org/10.1088/0264-9381/30/23/235007}{\emph{Class. Quant.
  Grav.} {\bf 30} (2013) 235007}, [\href{http://arxiv.org/abs/1307.6800}{{\tt
  1307.6800}}].

\bibitem{christodoulou2012formation}
D.~Christodoulou, \emph{The formation of black holes in general relativity},
  in \emph{The Twelfth Marcel Grossmann Meeting: On Recent Developments in
  Theoretical and Experimental General Relativity, Astrophysics and
  Relativistic Field Theories (In 3 Volumes)}, pp.~24--34, World Scientific,
  2012.

\bibitem{Bhattacharjee:2016zof}
S.~Bhattacharjee, S.~Sarkar and A.~Virmani, \emph{{Internal Structure of
  Charged AdS Black Holes}},
  \href{http://dx.doi.org/10.1103/PhysRevD.93.124029}{\emph{Phys. Rev.} {\bf
  D93} (2016) 124029}, [\href{http://arxiv.org/abs/1604.03730}{{\tt
  1604.03730}}].

\bibitem{Dafermos:2012np}
M.~Dafermos, \emph{{Black holes without spacelike singularities}},
  \href{http://dx.doi.org/10.1007/s00220-014-2063-4}{\emph{Commun. Math. Phys.}
  {\bf 332} (2014) 729--757}, [\href{http://arxiv.org/abs/1201.1797}{{\tt
  1201.1797}}].

\bibitem{Dafermos:2017dbw}
M.~Dafermos and J.~Luk, \emph{{The interior of dynamical vacuum black holes I:
  The $C^0$-stability of the Kerr Cauchy horizon}},
  \href{http://arxiv.org/abs/1710.01722}{{\tt 1710.01722}}.

\bibitem{dafermos2018rough}
M.~Dafermos and Y.~Shlapentokh-Rothman, \emph{Rough initial data and the
  strength of the blue-shift instability on cosmological black holes with
  $\lambda$> 0}, {\emph{Classical and Quantum Gravity} {\bf 35} (2018) 195010}.

\bibitem{Cardoso:2017soq}
V.~Cardoso, J.~L. Costa, K.~Destounis, P.~Hintz and A.~Jansen,
  \emph{{Quasinormal modes and Strong Cosmic Censorship}},
  \href{http://dx.doi.org/10.1103/PhysRevLett.120.031103}{\emph{Phys. Rev.
  Lett.} {\bf 120} (2018) 031103}, [\href{http://arxiv.org/abs/1711.10502}{{\tt
  1711.10502}}].

\bibitem{Hod:2018dpx}
S.~Hod, \emph{{Strong cosmic censorship in charged black-hole spacetimes: As
  strong as ever}},
  \href{http://dx.doi.org/10.1016/j.nuclphysb.2019.03.003}{\emph{Nucl. Phys.}
  {\bf B941} (2019) 636--645}, [\href{http://arxiv.org/abs/1801.07261}{{\tt
  1801.07261}}].

\bibitem{Dias:2018ufh}
O.~J.~C. Dias, H.~S. Reall and J.~E. Santos, \emph{{Strong cosmic censorship
  for charged de Sitter black holes with a charged scalar field}},
  \href{http://dx.doi.org/10.1088/1361-6382/aafcf2}{\emph{Class. Quant. Grav.}
  {\bf 36} (2019) 045005}, [\href{http://arxiv.org/abs/1808.04832}{{\tt
  1808.04832}}].

\bibitem{Mo:2018nnu}
Y.~Mo, Y.~Tian, B.~Wang, H.~Zhang and Z.~Zhong, \emph{{Strong cosmic censorship
  for the massless charged scalar field in the Reissner-Nordstrom–de Sitter
  spacetime}}, \href{http://dx.doi.org/10.1103/PhysRevD.98.124025}{\emph{Phys.
  Rev.} {\bf D98} (2018) 124025}, [\href{http://arxiv.org/abs/1808.03635}{{\tt
  1808.03635}}].

\bibitem{Hod:2018lmi}
S.~Hod, \emph{{Quasinormal modes and strong cosmic censorship in near-extremal
  Kerr–Newman–de Sitter black-hole spacetimes}},
  \href{http://dx.doi.org/10.1016/j.physletb.2018.03.020}{\emph{Phys. Lett.}
  {\bf B780} (2018) 221--226}, [\href{http://arxiv.org/abs/1803.05443}{{\tt
  1803.05443}}].

\bibitem{Luk:2017jxq}
J.~Luk and S.-J. Oh, \emph{{Strong cosmic censorship in spherical symmetry for
  two-ended asymptotically flat initial data I. The interior of the black hole
  region}},  \href{http://arxiv.org/abs/1702.05715}{{\tt 1702.05715}}.

\bibitem{Luna:2018jfk}
R.~Luna, M.~Zilhão, V.~Cardoso, J.~L. Costa and J.~Natário, \emph{{Strong
  Cosmic Censorship: the nonlinear story}},
  \href{http://dx.doi.org/10.1103/PhysRevD.99.064014}{\emph{Phys. Rev.} {\bf
  D99} (2019) 064014}, [\href{http://arxiv.org/abs/1810.00886}{{\tt
  1810.00886}}].

\bibitem{Dias:2018ynt}
O.~J.~C. Dias, F.~C. Eperon, H.~S. Reall and J.~E. Santos, \emph{{Strong cosmic
  censorship in de Sitter space}},
  \href{http://dx.doi.org/10.1103/PhysRevD.97.104060}{\emph{Phys. Rev.} {\bf
  D97} (2018) 104060}, [\href{http://arxiv.org/abs/1801.09694}{{\tt
  1801.09694}}].

\bibitem{Dias:2018etb}
O.~J.~C. Dias, H.~S. Reall and J.~E. Santos, \emph{{Strong cosmic censorship:
  taking the rough with the smooth}},
  \href{http://dx.doi.org/10.1007/JHEP10(2018)001}{\emph{JHEP} {\bf 10} (2018)
  001}, [\href{http://arxiv.org/abs/1808.02895}{{\tt 1808.02895}}].

\bibitem{hiscock1977stress}
W.~A. Hiscock, \emph{Stress-energy tensor near a charged, rotating, evaporating
  black hole}, {\emph{Physical Review D} {\bf 15} (1977) 3054}.

\bibitem{birrell1978falling}
N.~Birrell and P.~Davies, \emph{On falling through a black hole into another
  universe}, {\emph{Nature} {\bf 272} (1978) 35}.

\bibitem{Dias:2019ery}
O.~J.~C. Dias, H.~S. Reall and J.~E. Santos, \emph{{The BTZ black hole violates
  strong cosmic censorship}},  \href{http://arxiv.org/abs/1906.08265}{{\tt
  1906.08265}}.

\bibitem{Sela:2018xko}
O.~Sela, \emph{{Quantum effects near the Cauchy horizon of a
  Reissner-Nordström black hole}},
  \href{http://dx.doi.org/10.1103/PhysRevD.98.024025}{\emph{Phys. Rev.} {\bf
  D98} (2018) 024025}, [\href{http://arxiv.org/abs/1803.06747}{{\tt
  1803.06747}}].

\bibitem{Steif:1993zv}
A.~R. Steif, \emph{{The Quantum stress tensor in the three-dimensional black
  hole}}, \href{http://dx.doi.org/10.1103/PhysRevD.49.R585}{\emph{Phys. Rev.}
  {\bf D49} (1994) 585--589}, [\href{http://arxiv.org/abs/gr-qc/9308032}{{\tt
  gr-qc/9308032}}].

\bibitem{wald1994quantum}
R.~M. Wald, \emph{Quantum field theory in curved spacetime and black hole
  thermodynamics}.
\newblock University of Chicago Press, 1994.

\bibitem{Haag:1992hx}
R.~Haag, \emph{{Local quantum physics: Fields, particles, algebras, 2nd ed.}}
\newblock Springer, 1992.

\bibitem{birrell1984quantum}
N.~Birrell and P.~Davies, \emph{{Quantum fields in curved space}}.
\newblock Cambridge Univ Press, 1986.

\bibitem{Papadodimas:2013wnh}
K.~Papadodimas and S.~Raju, \emph{{Black Hole Interior in the Holographic
  Correspondence and the Information Paradox}},
  \href{http://dx.doi.org/10.1103/PhysRevLett.112.051301}{\emph{Phys. Rev.
  Lett.} {\bf 112} (2014) 051301}, [\href{http://arxiv.org/abs/1310.6334}{{\tt
  1310.6334}}].

\bibitem{Papadodimas:2013jku}
K.~Papadodimas and S.~Raju, \emph{{State-Dependent Bulk-Boundary Maps and Black
  Hole Complementarity}},
  \href{http://dx.doi.org/10.1103/PhysRevD.89.086010}{\emph{Phys. Rev.} {\bf
  D89} (2014) 086010}, [\href{http://arxiv.org/abs/1310.6335}{{\tt
  1310.6335}}].

\bibitem{Mathur:2011uj}
S.~D. Mathur, \emph{{What the information paradox is {\it not}}},
  \href{http://arxiv.org/abs/1108.0302}{{\tt 1108.0302}}.

\bibitem{Almheiri:2012rt}
A.~Almheiri, D.~Marolf, J.~Polchinski and J.~Sully, \emph{{Black Holes:
  Complementarity or Firewalls?}},
  \href{http://dx.doi.org/10.1007/JHEP02(2013)062}{\emph{JHEP} {\bf 02} (2013)
  062}, [\href{http://arxiv.org/abs/1207.3123}{{\tt 1207.3123}}].

\bibitem{Maldacena:1997re}
J.~M. Maldacena, \emph{{The Large N limit of superconformal field theories and
  supergravity}}, \href{http://dx.doi.org/10.1023/A:1026654312961}{\emph{Int.
  J. Theor. Phys.} {\bf 38} (1999) 1113--1133},
  [\href{http://arxiv.org/abs/hep-th/9711200}{{\tt hep-th/9711200}}].

\bibitem{Maldacena:2001kr}
J.~M. Maldacena, \emph{{Eternal black holes in anti-de Sitter}}, {\emph{JHEP}
  {\bf 0304} (2003) 021}, [\href{http://arxiv.org/abs/hep-th/0106112}{{\tt
  hep-th/0106112}}].

\bibitem{Raju:2018zpn}
S.~Raju, \emph{{A Toy Model of the Information Paradox in Empty Space}},
  \href{http://arxiv.org/abs/1809.10154}{{\tt 1809.10154}}.

\bibitem{Papadodimas:2015jra}
K.~Papadodimas and S.~Raju, \emph{{Remarks on the necessity and implications of
  state-dependence in the black hole interior}},
  \href{http://dx.doi.org/10.1103/PhysRevD.93.084049}{\emph{Phys. Rev.} {\bf
  D93} (2016) 084049}, [\href{http://arxiv.org/abs/1503.08825}{{\tt
  1503.08825}}].

\bibitem{Chamblin:1999tk}
A.~Chamblin, R.~Emparan, C.~V. Johnson and R.~C. Myers, \emph{{Charged AdS
  black holes and catastrophic holography}},
  \href{http://dx.doi.org/10.1103/PhysRevD.60.064018}{\emph{Phys. Rev.} {\bf
  D60} (1999) 064018}, [\href{http://arxiv.org/abs/hep-th/9902170}{{\tt
  hep-th/9902170}}].

\bibitem{hairer1993sod}
E.~Hairer, S.~N{\o}rsett and G.~Wanner, \emph{{Solving ordinary differential
  equations I: nonstiff problems}}.
\newblock Springer-Verlag, New York, 1993.

\bibitem{calcode1999}
C.~K. Raju, \emph{CALCODE: An ODE Solver}, 1999.

\bibitem{galassi2009gnu}
M.~Galassi, J.~Davies, J.~Theiler, B.~Gough, G.~Jungman, M.~Booth et~al.,
  \emph{GNU Scientific Library Reference Manual (Network Theory Ltd.)}, 2009.

\bibitem{Tange2011a}
O.~Tange, \emph{Gnu parallel - the command-line power tool}, {\emph{;login: The
  USENIX Magazine} {\bf 36} (Feb, 2011) 42--47}.

\bibitem{Lanir:2018vgb}
A.~Lanir, A.~Ori, N.~Zilberman, O.~Sela, A.~Maline and A.~Levi, \emph{{Analysis
  of quantum effects inside spherical charged black holes}},
  \href{http://dx.doi.org/10.1103/PhysRevD.99.061502}{\emph{Phys. Rev.} {\bf
  D99} (2019) 061502}, [\href{http://arxiv.org/abs/1811.03672}{{\tt
  1811.03672}}].

\bibitem{Balasubramanian:2004zu}
V.~Balasubramanian and T.~S. Levi, \emph{{Beyond the veil: Inner horizon
  instability and holography}},
  \href{http://dx.doi.org/10.1103/PhysRevD.70.106005}{\emph{Phys.Rev.} {\bf
  D70} (2004) 106005}, [\href{http://arxiv.org/abs/hep-th/0405048}{{\tt
  hep-th/0405048}}].

\bibitem{KeskiVakkuri:1998nw}
E.~Keski-Vakkuri, \emph{{Bulk and boundary dynamics in BTZ black holes}},
  \href{http://dx.doi.org/10.1103/PhysRevD.59.104001}{\emph{Phys. Rev.} {\bf
  D59} (1999) 104001}, [\href{http://arxiv.org/abs/hep-th/9808037}{{\tt
  hep-th/9808037}}].

\bibitem{Candelas:1980zt}
P.~Candelas, \emph{{Vacuum Polarization in Schwarzschild Space-Time}},
  \href{http://dx.doi.org/10.1103/PhysRevD.21.2185}{\emph{Phys. Rev.} {\bf D21}
  (1980) 2185--2202}.

\end{thebibliography}\endgroup

\end{document}